\documentclass{aa}  

\usepackage{graphicx}
\usepackage[colorlinks=true,     linkcolor=blue, citecolor=blue, filecolor=blue, urlcolor=blue]{hyperref}
\usepackage{txfonts}
\usepackage{xspace}
\usepackage{upgreek}
 \usepackage[dvipsnames]{xcolor}
\newcommand{\F}{\textit{Fermi}\xspace}
\newcommand{\hi}{H$\,\scriptstyle{\mathrm{I}}$\xspace}
\newcommand{\hii}{H$\,\scriptstyle{\mathrm{II}}$\xspace}

\newcommand{\gaussian}{Gaussian patch\xspace}
\newcommand{\outflow}{Source A\xspace}
\newcommand{\rightblob}{Source B\xspace}
\newcommand{\leftblob}{Source C\xspace}
\newcommand{\centerblob}{Source D\xspace}
\newcommand{\sect}{Sect.\xspace}
\newcommand{\fig}{Fig.\xspace}

\newcommand{\irii}[1]{#1}

\newcommand{\latrv}[1]{%
  \ifmmode
    {#1}%
  \else
    {#1}%
  \fi
}
\newcommand{\pbrv}[1]{%
  \ifmmode
    {{#1}}%
  \else
    {{#1}}%
  \fi
}

\newcommand{\jrone}[1]{%
  \ifmmode
    {#1}%
  \else
    {#1\xspace}%
  \fi
}

\begin{document}

   \title{Extended gamma-ray emission in the vicinity of the \object{Westerlund 1} massive star cluster and \object{Kes 41} supernova remnant seen by the \F Large Area Telescope}
   \titlerunning{\object{Westerlund 1} and \object{Kes 41} with \F LAT}
   
   \author{
   L. Tibaldo\inst{1}
   \and R. Bernet\inst{1,2}
   \and L. H\"arer\inst{3}
   \and M. Lemoine-Goumard\inst{4}
   \and L. Mohrmann\inst{3}
   \and G. Peron\inst{5} 
   \and B. Reville\inst{3}
   \and T. Vieu\inst{3}
   }
   \authorrunning{Tibaldo et al.}

  \institute{
  Univ Toulouse, CNES, CNRS, IRAP, Toulouse, France
  \and Institut Supérieur de l’Aéronautique et de l’Espace (ISAE-Supaero), Toulouse, France
  \and Max-Planck-Institut für Kernphysik, Heidelberg, Germany
  \and Université Bordeaux, CNRS, LP2I Bordeaux, UMR 5797, Gradignan, France
  \and INAF Osservatorio Astrofisico Arcetri, Florence, Italy
            }
%

\abstract
   {There is growing evidence for cosmic-ray acceleration in massive stellar clusters. Furthermore, extended gamma-ray emission suggests that particle transport in the vicinity of their sources is influenced by physical processes markedly different from large-scale diffusion in the Milky Way.}
   {We characterize extended gamma-ray emission in the direction of the \object{Westerlund 1} stellar cluster and \object{Kes 41} supernova remnant using $>16$ years of data from the \textit{Fermi} Large Area Telescope (LAT) at energies $>0.8$~GeV.}
   {We test whether clusters of gamma-ray sources not associated to multiwavelength counterparts are better described by extended emission components and characterize their spatial and spectral properties, as well as correlation with interstellar structures.}
   {We report the detection of three new extended emission components with soft spectra towards regions of high gas column density in the Galactic plane. \irii{One extended component associated with the natal cloud of \object{Kes 41} is statistically preferred over the point source previously reported towards the supernova remnant shell.} The other two \irii{extended components} overlap with neutral gas within $\sim$100~pc from the edge of the \object{Westerlund 1} superbubble. The extended emission may be explained either by mismodeled gas in the interstellar background model or by the local injection of particles. Under the latter hypothesis, explaining the component associated with \object{Kes 41} requires converting $\lesssim 5\%$ of the supernova remnant energy into accelerated particles, while accounting for the two components near \object{Westerlund 1} requires converting $10^{-4}$ of the cluster wind mechanical power into gamma rays.}
   {More detailed physical modeling and more accurate observational constraints are needed to conclude whether the new extended emission components are due to particles escap\latrv{ed} from \object{Westerlund 1} and \object{Kes 41}. Nevertheless, our results strengthen the evidence for gamma-ray emission structures arising at intermediate spatial scales between isolated objects and the large-scale diffuse emission from the interstellar medium. This could explain a part of the soft unassociated Galactic sources detected by the \F~LAT.}


   \maketitle
   \nolinenumbers
%

\section{Introduction}

The standard paradigm for cosmic rays (CRs) in the energy range from GeV to PeV is based on the two hypotheses that they are accelerated in the disk of the Milky Way, presumably in supernova remnants (SNRs), and then they are diffusively confined in a kpc-sized Galactic halo for durations of several Myr \citep{ginzburgOriginCosmic1964}. Recently, conventional models of Galactic CRs based on this paradigm have been called into question by the exquisite detail reached by observations and by difficulties in connecting the phenomenological description of the macro-observables to the micro-physics of particle acceleration and transport \citep[e.g.,][and references therein]{gabiciOriginGalactic2019}. Most notably, the standard paradigm is challenged by the low rate of SNR PeVatrons suggested both by theory and observations \citep[e.g.,][]{cristofariLowRate2020}, which is related to the problem of achieving a sufficient amplification of the magnetic field possibly via CR induced plasma instabilities \citep[e.g.,][]{Bampl2008}.  

Star-forming regions and massive stellar clusters (MSCs) have been invoked as possible additional sources of Galactic CRs to mitigate open issues of the standard paradigm. In addition to hosting many classes of energetic objects at the endpoint of massive star evolution, including SNRs, star-forming regions could provide specific sites and mechanisms of particle acceleration due to the concentration of multiple objects in a small region of space: termination shocks from the collective winds of MSCs \citep[e.g.,][]{morlinoParticleAcceleration2021} and the combination of multiple shocks from stellar winds and SNRs in superbubbles \citep[e.g.,][]{vieuCosmicRay2022}. Collective wind outflows from MSCs could play a role in the acceleration of particles up to multi-TeV energies \cite[e.g.][]{vieuMassiveStarCluster2023}. Furthermore, material injected by the winds of Wolf-Rayet stars enriched in He-burning products could help to explain the isotopic abundances observed in CRs, notably $^{22}$Ne/$^{20}$Ne, even though the exact mechanisms by which the massive-star material is incorporated into CRs and the implications for the acceleration site/process are still subject to debate \citep[e.g.,][]{rauch2009,lingenfelterOriginCosmic2019,bykovHighEnergyParticles2020,tatischeff2021}.

The standard Galactic CR paradigm also overlooks the physical processes by which particles escape their acceleration sites and merge with the large-scale CR population \citep[e.g.,][]{marcowith2025}.  Star-forming regions and, oftentimes, the vicinities of SNRs are characterized by large gas and radiation field densities as well as high levels of turbulence from winds and supernova explosions and, possibly, from the streaming of CRs themselves. Thus, amplified energy losses and lower diffusivities may give rise to an effective confinement of the particles around their acceleration sites and play a significant role in the CR lifecycle.

Interactions of CRs with interstellar matter produce gamma rays through hadronic processes (mostly creation and decay of neutral pions) as well as lepton Bremsstrahlung, in addition to inverse-Compton (IC) radiation from the scattering of CR leptons off low-energy photons from radio to optical. Therefore,  gamma-ray emission is a probe of CR acceleration and transport, see, e.g., \citet{strongCosmicRayPropagation2007,grenierNineLivesCosmic2015,tibaldoGammaRays2021}. Interestingly, over the past decade, both space-born and ground-based gamma-ray telescopes provided detections of extended gamma-ray sources in the direction of several star-forming regions and SNRs \citep[see, e.g., the compilation in][]{tibaldoGammaRays2021}.

However, the interpretation of the extended gamma-ray emission poses several challenges due to its coincidence with complex regions in the Galactic plane. The putative signal from particles accelerated in situ needs to be disentangled from the background due to the large-scale population of Galactic CRs, and results are sensitive to the modeling of the latter at GeV energies  \citep[e.g.,][]{sahaMorphologicalSpectral2020}. Furthermore, owing to the limited angular resolution of gamma-ray telescopes, it is usually difficult to clearly pinpoint the exact origin of the gamma-ray emission. The confusion between backgrounds, individual emitters, and collective phenomena makes it challenging to identify the specific acceleration and transport mechanisms that may be at play \citep[see, e.g., the case of the Cygnus X region discussed in][]{astiasarainMultipleEmission2023}. 

\object{Westerlund 1} is the most massive young stellar cluster known in the Milky Way with a mass of the order of $10^5\; M_\sun$ and stellar properties which suggest an approximate age of 4~Myr  \citep{clarkMassiveStellarPopulation2005,brandnerIntermediateLowmassStellar2008}, with possible subpopulations formed over several Myr and maximum ages reaching 10 Myr \citep{beasorAgeWesterlund12021,navareteDistanceAgeMassive2022}. Independent analyses of \textit{Gaia} EDR3 parallaxes based on careful selections of cluster members agree on a distance of $\sim$4~kpc: $4.23^{+0.23}_{-0.21}$~kpc according to \citet{negueruelaWesterlund1Light2022},  $4.06^{+0.36}_{-0.34}$~kpc according to \citet{navareteDistanceAgeMassive2022}. \latrv{The cluster is very compact, with a major semi-axis of $3.5'$ according to \citet{negueruelaWesterlund1Light2022}}. Based on th\latrv{e aforementioned} distance, the estimates of the star wind velocities and mass loss rates \citep{munoDiffuseNonthermalXRay2006,kavanaghDiffuseThermalXray2011,fenechALMA3Mm2018} make it possible to assess the mechanical wind luminosity of the cluster to $\lesssim10^{39}$~erg~s$^{-1}$.

Gamma-ray emission in the vicinity of \object{Westerlund 1} was detected at TeV \citep{abramowskiDiscoveryExtendedVHE2012} and GeV energies \citep{ohmGrayEmissionWesterlund2013}. Deep H.E.S.S. observations revealed gamma-ray emission above 0.37~TeV around the MSC with a ring-like morphology extending over $\sim$2\degr\xspace \citep{aharonianDeepSpectromorphologicalStudy2022}. Interestingly, the distance of the gamma-ray ring from the MSC is compatible with the expected location of the collective wind termination shock. \citet{harerUnderstandingTeVGray2023} presented a model of particle acceleration at the MSC wind termination shock capable of reproducing H.E.S.S. observations in which the bulk of the gamma-ray emission is produced by IC scattering.
\citet{lemoinegoumardFermiWd12025} extended the characterization of the gamma-ray emission down to 3~GeV using data from the \F~Large Area Telescope (LAT). The peak of the emission is offset with respect to TeV, but connects to it smoothly spectrally and spatially. The peak of the emission above 3~GeV coincides with a low-density cavity in the interstellar medium (ISM) towards the outskirt of the Galactic disk, which lead \citet{lemoinegoumardFermiWd12025} to argue that it originates via IC scattering in a nascent CR-loaded outflow.

About 2\degr\ from the line of sight of \object{Westerlund 1} lies SNR \object{Kes 41}, also known as G337.8-00.1. 
\object{Kes 41} is detected in radio as an elongated shell with a size of $9' \times 6'$ \citep{whiteoakMOSTSupernovaRemnant1996}. Interaction between the SNR and dense molecular material is supported by the detection of the OH maser 1720 MHz line 
\citep{koraleskyShockexcitedMaserEmission1998}. This hypothesis is strengthened by X-ray measurements which reveal \latrv{thermal emission from the central region of the SNR} \citep[and no compact source,][]{combiDiscoveryThermalXray2008}. X-ray data point to a high foreground column density $> 6.9 \times 10^{22}$~cm$^{-2}$ which suggests a relatively distant location ($\ge7$ kpc), placing the SNR at the far kinematic distance of $\sim$12~kpc. \latrv{\citet{zhangMetalenrichedThermalComposite2015} show that evolutionary scenarios in which the SNR evolved in a uniform or clumpy medium are not consistent with the observations. They conclude that \object{Kes 41} is the result of a supernova explosion in a pre-existing cavity, and that the SNR has left the adiabatic stage of evolution and entered the radiative phase after the shock encountered the dense molecular gas beyond the cavity. The ionization inferred from the X-ray spectral analysis implies an age of $4\; \mathrm{to}\; 100 f^{1/2}$~kyr (with $f$ filling factor of the ionized gas).} Gamma-ray emission from \object{Kes 41} was reported in \citet{hewittCorrelationSupernovaRemnant2009}. Association over 1\degr\ between CO emission in the MOPRA survey and gamma-ray emission measured by the LAT  was discussed in \citet{supanUnidentifiedGrayEmission2018,supanNatalMolecularCloud2018}, who identify the cloud as the birthsite of \object{Kes 41}.

This work aims at extending the analysis of \F~LAT data in the vicinity of \object{Westerlund 1} and \object{Kes 41} to lower energies with respect to \citet{lemoinegoumardFermiWd12025}, with a particular focus on unassociated sources reported in LAT catalogs, extended gamma-ray emission and the structures of the ISM in this region. Section~\ref{sec:data} presents the gamma-ray dataset and general analysis framework. Section~\ref{sec:analysis} describes the analysis of the data and the main results. Section~\ref{sec:discussion} discusses the new extended emission components found in our analysis. Finally, \pbrv{\sect}~\ref{sec:conclusions} provides a summary of the results and of our conclusions.

\section{Dataset and analysis framework}\label{sec:data}

The LAT is the main instrument on the \textit{Fermi Gamma-ray Space Telescope} \citep{atwoodLARGEAREA2009}, detecting gamma rays via pair production. We use $\sim$16.4 years\footnote{From the beginning of LAT scientific operations, on 2008-08-04 15:43:36, to 2024-12-25 00:00:00.} of LAT \latrv{Pass 8 (P8R3) data} \citep{atwoodPass8Full2013,bruelFermiLATImprovedPass82018}. We analyze the data using the dedicated analysis software \texttt{fermitools}\footnote{\url{https://fermi.gsfc.nasa.gov/ssc/data/analysis/documentation/}} version 2.2.0 and the Python package \texttt{fermipy} \citep{fermipy} version 1.3.1.

We select \latrv{events} belonging to the \texttt{CLEAN} class that provides good sensitivity for analysis of point sources and moderately extended sources and has low background at higher energies. We apply the standard data quality filters \texttt{(DATA\_QUAL==1 \&\& LAT\_CONFIG==1)}.

\begin{table}
\caption{The four datasets used in the analysis.}             
\label{tab:esel}      
\centering                          
\begin{tabular}{cc}        
\hline\hline                 
PSF type & minimum energy (GeV)  \\    
\hline                        
PSF0 & 10 \\
PSF1 & 2 \\
PSF2 & 1 \\
PSF3 & 0.8 \\
\hline                                   
\end{tabular}
\end{table}

We separate the candidate photons in four independent datasets depending on their PSF event type, that is, the quality of the direction reconstruction. As detailed in \pbrv{Table}~\ref{tab:esel}, for each event type we set a lower energy threshold such that we have always a PSF~68\% containment radius better than 0.5\degr, which roughly corresponds to the expected angular size of the \object{Westerlund 1} wind termination shock and to the observed angular size of the main ISM structures in the region. The resulting minimum energy threshold for the analysis is 0.8 GeV. The maximum energy is 1~TeV for all event types. For the four event types we select events with a measured zenith angle $< 105\degr$, which we verified to be sufficient to exclude bright emission from CR interactions in the Earth's atmosphere.

We analyze the four datasets using a joint binned maximum likelihood method. We bin candidate photons over a region of interest (ROI) in Galactic coordinates spanning $13.8\degr \times 18\degr$ and centered at $l = 337.45\degr$, $b = -0.40\degr$. The ROI is not centered on \object{Westerlund 1} or \object{Kes 41} so that we \latrv{exclude the region around the bright SNR RX~J1713.7$-$3946 \citep[that has a complex morphology, see][]{abdallaHESSObservationsRX2018} and we prevent potential residuals from interfering with the fit of the other components}, but we have an ROI large enough to effectively separate the different components in our model (see the description below).  We bin the events on a grid with 0.1\degr\ angular step on the sky and ten bins per decade in energy. The LAT energy dispersion is taken into account with \texttt{edisp\_bins=$-$2}.

Models for the ROI are compared and selected based on their likelihood, number of degrees of freedom (d.o.f.), and quality of the data-model residuals. For nested models we use the likelihood ratio test based on the quantity $\mathrm{TS} = 2 (\ln \mathcal{L}_1 - \ln \mathcal{L}_0)$, which is distributed as a $\chi^2$ with $(n_1 - n_0)$ degrees of freedom \citep[e.g.,][]{protassovStatisticsHandleCare2002}. For non-nested models we employ the Akaike information criterion $\mathrm{AIC} = 2 [(n_1 - n_0) - (\ln \mathcal{L}_1 - \ln \mathcal{L}_0)]$. Negative AIC values indicate that model 1 provides a better representation of the data at a smaller cost in terms of free parameters according to information theory  \citep[e.g.,][]{burnhamModelSelectionMultimodel2004}. The quality of the data--model residuals is assessed using the so-called ``PS map'' \citep{bruelNewMethodPerform2021}. At every step of the analysis we check the PS maps and verify that the model selected has similar or smaller residuals with respect to the one it is being compared to.

   \begin{figure}
   \centering
   \includegraphics[width=\hsize]{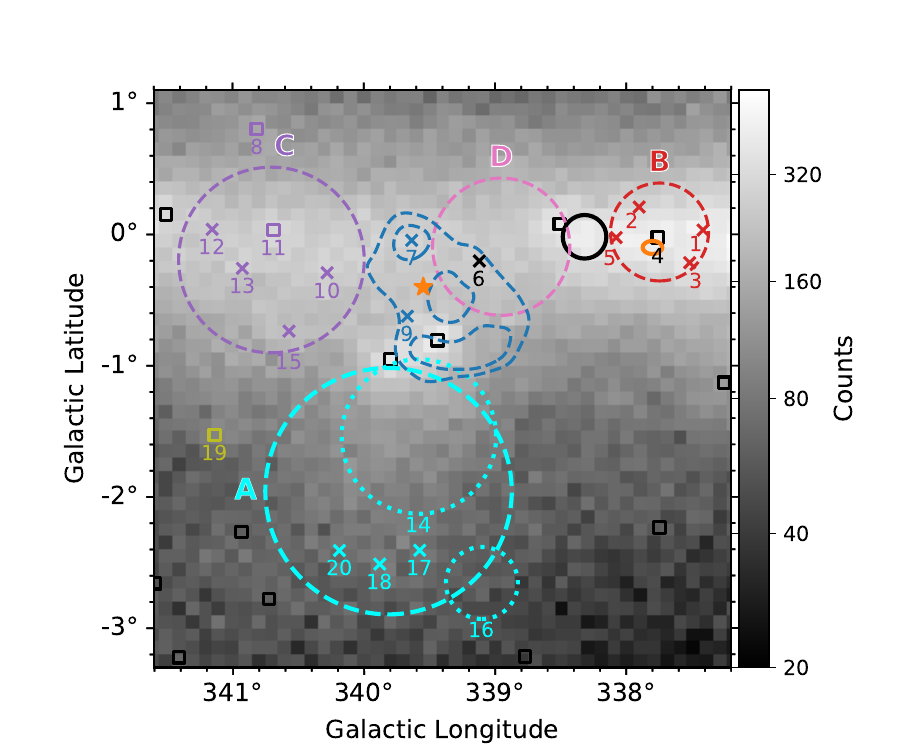}
      \caption{4FGL-DR4 sources in the region around \object{Westerlund 1} and \object{Kes 41} \citep{collaborationIncrementalFermiLarge2022,balletFermiLargeArea2024}. The orange star marks the position of \object{Westerlund 1} \citep{tarricq3DKinematicsAge2021}, while the orange ellipse corresponds to the radio shell of SNR G337.8$-$0.1, alias \object{Kes 41} \citep{greenUpdatedCatalogue3102025}. Blue dashed contours correspond to TeV gamma-ray emission measured with H.E.S.S. and attributed to \object{Westerlund 1} \citep{aharonianDeepSpectromorphologicalStudy2022}.
      Squares/crosses mark the positions of point-like sources and continuous/dotted circles indicate the 68\% containment  regions of extended sources in 4FGL-DR4 for sources that are kept/discarded in the model in \pbrv{\sect}~\ref{sec:geomodel}. Unassociated and unknown type sources within 2.5\degr\ from \object{Westerlund 1} are highlighted in color, with color coding that reflects the extended emission regions discussed in the paper.  Sources discussed individually in the text are numbered, and \pbrv{Table}~\ref{tab:unassoc} provides the correspondence of number to 4FGL-DR4 name.
      Dashed circles correspond to the  68\% containment regions of extended sources added in the model in \pbrv{\sect}~\ref{sec:geomodel}. The background map shows LAT counts above 800~MeV for the dataset used in our analysis.
              }
         \label{fig:mapunassoc}
   \end{figure}

\section{Analysis and results}\label{sec:analysis}

\subsection{Sky model}

We start from the sky model in the fourth data release of the fourth catalog of LAT gamma-ray source \citep[4FGL-DR4][]{collaborationIncrementalFermiLarge2022,balletFermiLargeArea2024}. The sky model \pbrv{ consists of individual sources and background models\footnote{\url{https://fermi.gsfc.nasa.gov/ssc/data/access/lat/BackgroundModels.html}}, including} interstellar emission from the interactions of the large-scale population of Galactic CRs with the ISM and isotropic background spectral templates (specific to each event type) accounting for extragalactic diffuse emission and residual contamination from charged CRs misclassified as gamma rays.
Figure~\ref{fig:mapunassoc} shows the 4FGL-DR4 sources in the region around \object{Westerlund 1} \citep{collaborationIncrementalFermiLarge2022,balletFermiLargeArea2024}. Within 2.5 \degr\ from \object{Westerlund 1} we count 18 sources which lack an association to an object that belongs to a class of known gamma-ray emitters from multi-wavelength catalogs.

The treatment of the interstellar background in our analysis and some preliminary adjustments of the sky model are described in \pbrv{Appendix}~\ref{app:prelfit}. This results in the following modifications to the 4FGL-DR4 model:
\begin{itemize}
\item the ``patch'' component in the interstellar background model (diffuse emission of unknown origin that is iteratively determined from residuals in the LAT data) is replaced by an extended source with 2D Gaussian morphology, hereafter the \gaussian;
\item following \citet{lemoinegoumardFermiWd12025}, three 4FGL-DR4 sources (the sources numbered as 6, 7, and 9 in \pbrv{\fig}~\ref{fig:mapunassoc}) are replaced by two templates modeling the ring-like feature around \object{Westerlund 1} detected by H.E.S.S. \citep{aharonianDeepSpectromorphologicalStudy2022};  
\item five unassociated 4FGL-DR4 sources (the sources numbered as 14, 16, 17, 18, and 20 in \pbrv{\fig}~\ref{fig:mapunassoc}) are replaced by an extended source with 2D Gaussian morphology, hereafter \outflow, modeling the tentative outflow from \object{Westerlund 1} discussed in \citet{lemoinegoumardFermiWd12025}.
\end{itemize}

After these preliminary model adjustments, we are left with 11 unassociated pointlike sources within 2.5\degr\ from \object{Westerlund 1}. There are two remarkable clusters: four sources at $l$ from $\sim$337\degr\ to $\sim$338\degr\ (highlighted in red in \pbrv{\fig}~\ref{fig:mapunassoc}) surrounding the source 4FGL~J1638.5$-$4657c (numbered as 4 in \pbrv{\fig}~\ref{fig:mapunassoc}) associated with the radio shell of the \object{Kes 41} SNR, and five sources plus one separated by $<1\degr$ at $l$ from $\sim$340\degr\ to $\sim$341\degr (highlighted in purple and numbered as 10, 11, 12, 13, and 15 plus 8 in \pbrv{\fig}~\ref{fig:mapunassoc}). \jrone{All these sources are modelled in 4FGL-DR4 using a curved log-parabola spectrum.} In this section we seek an improved model of the region by testing whether \jrone{they} can be replaced by extended emission components possibly associated to ISM structures.

\subsection{Morphological characterization with Gaussian/disk models}\label{sec:geomodel}
We first consider replacing clusters of unassociated sources with extended Gaussians or disks. For all new extended sources added in this section we use a log-parabola spectral model which can capture spectral curvature if present. Spectral optimization will be discussed later.

We start with the brighter cluster at $l$ from $\sim$337\degr\ to $\sim$338\degr . We replace the four unassociated sources 4FGL~J1636.9$-$4710c, 4FGL~J1638.1$-$4641c, 4FGL~J1638.4$-$4715c, and 4FGL~J1639.8$-$4642c  (numbered as 1, 2, 3, and 5 in \pbrv{\fig}~\ref{fig:mapunassoc}) with either a disk or a 2D Gaussian. We also test analogous models in which the central source 4FGL~J1638.5$-$4657c (numbered as 4 in \pbrv{\fig}~\ref{fig:mapunassoc}) associated with the \object{Kes 41} SNR is removed from the sky model. For this part of the analysis the normalization and spectral parameters of the latter, as well as the normalization of nearby extended sources (4FGL~J1633.0$-$4746e, 4FGL~J1631.6$-$4756e, and 4FGL~J1636.3$-$4731e) is kept free.

\begin{table}
\caption{Summary of the comparison between morphological models for the region of \rightblob  in \pbrv{\sect}~\ref{sec:geomodel}. The model selected is marked in bold. $-$5S corresponds to removing the five point sources numbered as 1, 2, 3, 4, and 5 in \pbrv{\fig}~\ref{fig:mapunassoc}. $-$4S to removing the four point sources 1, 2, 3, and 5.}             
\label{tab:geo_models_b}      
\centering                          
\begin{tabular}{l c c c}        
\hline\hline     
Model & $\Delta \ln \mathcal{L}$ & $\Delta$ d.o.f. & AIC\\
\hline
\bfseries{$-$4S + disk} & 12.9 & $-6$ & $-37.8$ \\ 
$-$4S + Gauss & 9.2 & $-6$ & $-30.4$ \\
$-$5S + disk & 2.0 & $-9$ & $-22.0$\\
$-$5S + Gauss & 6.5 & $-9$ & $-31.0$\\
\hline
\end{tabular}
\end{table}

Table~\ref{tab:geo_models_b} summarizes the results. The model favored with the most negative AIC, $-37.8$, is the one where we replace the four unassociated sources with a disk. In this model the central point-like source 4FGL~J1638.5$-$4657c \latrv{(source 4 in \pbrv{\fig}~\ref{fig:mapunassoc})} remains borderline significant, $\mathrm{TS}=21.8$, although below the formal threshold for inclusion in LAT catalogs of $\mathrm{TS}=25$. From now on this model becomes our baseline, and we refer to the disk source as \rightblob.

\begin{table}
\caption{Summary of the comparison between morphological models for the region of \leftblob  in \pbrv{\sect}~\ref{sec:geomodel}. The model selected is marked in bold. $-$5S corresponds to removing the five point sources numbered as 10, 11, 12, 13, and 15 in \pbrv{\fig}~\ref{fig:mapunassoc}. $-$4S to removing the four point sources 10, 12, 13, and 15.}             
\label{tab:geo_models_c}      
\centering                          
\begin{tabular}{l c c c}        
\hline\hline     
Model & $\Delta \ln \mathcal{L}$ & $\Delta$ d.o.f. & AIC\\
\hline
$-$5S + disk & $-21.0$ & $-9$ & 24.0 \\
$-$5S + Gauss & $-21.9$ & $-9$ & 25.8 \\
\bfseries{$-$4S + disk} & 7.8 & $-6$ & $-27.6$ \\
$-$4S + Gauss & 0.0 & $-6$ & $-12.0$\\
\hline
\end{tabular}
\end{table}

We then consider the cluster of unassociated sources at $l$ from $\sim$340\degr\ to $\sim$341\degr. We replace the five sources 4FGL~J1649.2$-$4513c, 4FGL~J1649.3$-$4441, 4FGL~J1650.9$-$4420c, 4FGL~J1651.4$-$4442c, and 4FGL~J1652.2$-$4516 (numbered as 10, 11, 12, 13, and 15 in \pbrv{\fig}~\ref{fig:mapunassoc}) with either a disk or a 2D Gaussian. For this part of the analysis we also keep free the normalization of two nearby unassociated sources 4FGL~J1646.5$-$4406 and 4FGL~J1657.7$-$4520 (numbered as 8 and 19, respectively, in \pbrv{\fig}~\ref{fig:mapunassoc}). These models are not favored based on AIC (\pbrv{Table}~\ref{tab:geo_models_c}) and they result in significant residuals towards the position of 4FGL~J1649.3$-$4441 (source 11 in \pbrv{\fig}~\ref{fig:mapunassoc}). Thus we also test models where the latter is kept in the model along with the disk or 2D Gaussian replacing the other four sources.

As demonstrated in \pbrv{Table}~\ref{tab:geo_models_c}, the favored model, $\mathrm{AIC} = -27.6$, is the one with four unassociated sources replaced by a disk and 4FGL~J1649.3$-$4441 kept in the model (with $\mathrm{TS}=36.6$). The two other unassociated sources nearby also remain significant. From now on this model becomes our baseline, and we refer to the new disk source as \leftblob.

After these model modifications we note the presence of significant residuals at $l \sim339\degr$, $\jrone{b \sim -0.1 \degr}$. We test models in which we add an extended source at this position, modeled by either a disk or a 2D Gaussian. We obtain a marked increase in log-likelihood, that is, the newly added source is significant. The best likelihood is obtained for the disk model, which gives $\mathrm{TS}=69$ for 6 additional degrees of freedom, that is a significance of $7.2\sigma$. Additionally, the source is significantly extended, with a TS$_\mathrm{ext}$ (TS comparing the disk model with a point-like model) of 53.8. From now on we include this extended source in our model, and we refer to it as \centerblob.

The model obtained at this point produces satisfactory residuals. However, the extended sources were fit one by one, which may influence the results. To test the robustness of the extended source models we iteratively refit the positions and extensions of the \gaussian, three extended sources with multiwavelength counterparts within 3\degr\ from \object{Westerlund 1}  the morphology of which is adjusted to the LAT data (4FGL~J1633.0$-$4746e, 4FGL~J1631.6$-$4756e, and 4FGL~J1636.3$-$4731e), \outflow, \rightblob, \leftblob, and \centerblob. We stop the iterative fit after three iterations  with a global  $\Delta \ln \mathcal{L} = 113.8$ and the third iteration resulting in $\Delta \ln \mathcal{L} < 1$.

\begin{table*}
    \caption{Morphological description of the extended sources in \pbrv{\sect}~\ref{sec:geomodel}. The first uncertainties are statistical, while the second are systematics from the variations of the interstellar background model discussed in \pbrv{Appendix}~\ref{app:sysunc}.}             
\label{tab:geo_models_results_morph}      
\centering                          
\begin{tabular}{l c c c c}        
\hline\hline     
Source & Spatial model & $l$ (\degr) & $b$ (\degr) & $R_{68}$ (\degr) \\
\hline
\outflow &  Gaussian & $339.81 \pm 0.05 ^{+0.00} _{-0.12}$ & $-1.96 \pm 0.05 ^{+0.05} _{-0.01}$ & $0.94^{+0.04} _{-0.02} \phantom{0}^{+0.07} _{-0.01}$ \\
\rightblob & Disk & $337.77\pm0.07 ^{+0.03}_{\latrv{-0.01}} $ & $0.02 \pm 0.04 ^{\latrv{+0.01}}_{-0.04}$ & $0.339^{+0.012}_{-0.012} \phantom{0}^{+0.12}_{-0.15}$ \\
\leftblob &  Disk & $340.71\pm0.06^{+0.02}_{-0.05}$ & $-0.20 \pm 0.06^{+0.03}_{-0.04}$ & $0.65^{+0.02}_{-0.02} \phantom{0}^{+0.00}_{-0.10}$ \\ 
\centerblob\tablefootmark{*} & Disk & $338.96\pm0.06^{+0.10}_{\latrv{-0.03}}$ & $-0.09 \pm 0.06 ^{+0.00}_{-\latrv{0.04}}$ & $0.48^{+0.04}_{-0.04} \phantom{0}^{+0.03}_{-0.04}$ \\
\hline
\end{tabular}
\tablefoot{
\tablefoottext{*}{The extension fit of \centerblob does not converge in a model variation with the 4FGL patch, thus the corresponding parameter uncertainties in the table do not include such model variation.}
}
\end{table*}

The best-fit morphological parameters after the iterative fits are provided in \pbrv{Table}~\ref{tab:geo_models_results_morph}. The table includes systematic uncertainties from a few variations of the interstellar emission model as described in \pbrv{Appendix}~\ref{app:sysunc}. We note that the extension fit of \centerblob does not converge in a model variation with the \gaussian replaced by the 4FGL patch, which contains structured emission in this region. \centerblob remains significant ($\mathrm{TS}=57.8$) even in this case, though, if the morphology is fixed to the determination with our baseline analysis (\pbrv{Table}~\ref{tab:geo_models_results_morph}). This suggests that the properties of this component should be regarded as particularly uncertain.

\begin{table*}
\caption{Summary of the comparison between spectral models for the extended emission components studied  in \pbrv{\sect}~\ref{sec:geomodel}. For each component we compare four spectral models: power law (PL), log-parabola (LP), power law with exponential cutoff (PLC), and smooth broken power law (BPL). Some values are missing in case of models for which convergence could not be achieved.}             
\label{table:spectral_models}      
\centering                          
\begin{tabular}{l c c c c}        
\hline\hline     
 Source & $\Delta \log \mathcal{L}_\mathrm{LP-PL}$ & $\Delta \log \mathcal{L}_\mathrm{PLC-PL}$ & $\Delta \log \mathcal{L}_\mathrm{BPL-PL}$ & Model selected \\
\hline
\gaussian &  17.9 & 11.9 & 10.5 & LP\\
\outflow &  0.6 &  -- & 0. & PL \\
\rightblob & 12.0 & 2.5 & 8.5 & LP \\
\leftblob & 16.0 & 15.9 & -- & LP \\ 
HESS 1 & 0.1 & -- & 0.0 & PL \\
\centerblob & 7.0 & 6.7 & -- & LP \\
HESS 2 & 0.0 & -- & 0.0 & PL\\
\hline                                   
\end{tabular}
\end{table*}

So far the extended emission components were modeled using power-law or log-parabola spectral functions based on previous results in the literature or educated guesses made for the newly added sources. For the \gaussian, the two templates based on H.E.S.S. data and other extended sources around \object{Westerlund 1} we test multiple spectral models (in order of decreasing source TS). The models considered\footnote{See \url{https://fermi.gsfc.nasa.gov/ssc/data/analysis/scitools/source_models.html} for the model description.} are: power law (2 degrees of freedom), log-parabola (3 degrees of freedom), power law with exponential cutoff (\texttt{PLSuperExpCutoff2}, 3 degrees of freedom), smooth broken power law (four degrees of freedom, with smoothing parameter fixed to 0.2). If $\mathrm{TS}_\mathrm{curv} = 2 \times \Delta \log \mathcal{L}$ between any models with curvature and the simple power law is $> 9$ we select the curved model with the most negative AIC, otherwise we use a simple power law. The results are provided in \pbrv{Table}~\ref{table:spectral_models}.

\begin{table*}
    \caption{Spectral description of the extended sources studied in \pbrv{\sect}~\ref{sec:geomodel}. The spectral model is parametrized as $\mathrm{d}N/\mathrm{d}E = K (E/E_0)^{-\alpha - \beta \ln (E/E_0)}$, therefore $\beta=0$ corresponds to a power-law spectrum and $\beta > 0$ to a curved log-parabola spectrum. We fix $E_0 = 1$~GeV for all sources. For log-parabola spectra we also provide the peak energy of $\jrone{E^2 \mathrm{d}N/\mathrm{d}E}$ \citep{collaborationIncrementalFermiLarge2022}. The first uncertainties are statistical, while the second are \jrone{systematic} from the variations of the LAT effective area and interstellar background model  discussed in \pbrv{Appendix}~\ref{app:sysunc}. For the latter, the morphological parameters are fixed to the reference values in \pbrv{Table}~\ref{tab:geo_models_results_morph}.}             
\label{tab:geo_models_results_spec}      
\centering                          
\begin{tabular}{l c c c  c c c}        
\hline\hline     
Source  & $K$ (cm$^{-2}$ s$^{-1}$ MeV$^{-1}$) & $\alpha$ & $\beta$ & $E_\mathrm{peak}$ (GeV) & TS$_\mathrm{curv}$\\
\hline
\outflow & $(1.63\pm 0.07^{+0.12}_{-0.19}) \times 10^{-11}$  & $1.99\pm 0.02^{+0.03}_{-0.03}$ & 0 & - & 1.2 \\
\rightblob & $(2.36\pm 0.10^{+0.44}_{-0.38}) \times 10^{-11}$  & $2.31\pm 0.06^{\latrv{+0.08}}_{-0.02}$ & $0.10\pm0.02^{+0.04}_{-0.01}$ & $0.2 \pm 0.2^{+0.1}_{\latrv{-0.1}}$ & 24.0\\
\leftblob & $(1.85\pm 0.13^{+0.74}_{-0.19}) \times 10^{-11}$  & $1.85\pm 0.13^{\latrv{+0.37}}_{-0.01}$ & $0.49\pm0.09^{+0.14}_{\latrv{-0.25}}$ & $1.16 \pm 0.24^{+0.02}_{\latrv{-0.52}}$ & 32.0 \\ 
\centerblob & $(6.7\pm 1.1^{+2.4}_{-2.7}) \times 10^{-12}$ & $1.0\pm 0.2^{\latrv{+0.78}}_{-0.56}$ & $1.05\pm0.15^{+0.8}_{-0.26}$ & $1.6 \pm 0.3^{+0.0}_{\latrv{-0.4}}$ & 14\\
\hline
\end{tabular}
\end{table*}

The best-fit spectral model parameters for the extended emission components are provided in \pbrv{Table}~\ref{tab:geo_models_results_spec}. The various extended emission components in the region are characterized by different spectral shapes, as shown by the preference for power law or log-parabola spectra and by the values of the spectral parameters, as well as by the gamma-ray maps in energy bands in \pbrv{Appendix}~\ref{app:maps_ebands}. 

\subsection{Correlation with ISM structures}\label{sec:ismmodel}

The H.E.S.S. ring-like feature and \outflow, characterized by power-law spectra, are located in regions of low ISM column densities as discussed in \citet{aharonianDeepSpectromorphologicalStudy2022,harerUnderstandingTeVGray2023,lemoinegoumardFermiWd12025}. Conversely, the three new extended sources introduced in our analysis with curved spectra are situated in the Galactic plane. Therefore, we explore the correlation between our gamma-ray sources and ISM structures. 

 As shown in \pbrv{Appendix}~\ref{app:covbands}, there are conspicuous CO features (total $W_\mathrm{CO} \gtrsim 100$~K~km~s$^{-1}$) towards \rightblob and \leftblob in the kinematic range from  $V_\mathrm{LSR} = -70$~km~s$^{-1}$ to $-40$~km~s$^{-1}$. Such velocities are consistent with those of kinematic tracers associated with \object{Kes 41} \citep[e.g.,][]{supanNatalMolecularCloud2018} and \object{Westerlund 1} \citep[][and references therein]{lemoinegoumardFermiWd12025}. \centerblob is overlapping with a CO emission feature in the kinematic range from $V_\mathrm{LSR} = -130$~km~s$^{-1}$ to $-100$~km~s$^{-1}$ with total $W_\mathrm{CO} \gtrsim 50$~K~km~s$^{-1}$.


To study in more detail the correlation between extended gamma-ray emission and interstellar gas we replace the disks with gas templates. The templates trace gas already included in the interstellar background model, therefore the gamma-ray emission attributed to the additional templates corresponds to excess emission on top of the background model. Excess emission can be due to a localized underestimation in the backgroun\latrv{d} model of either the gas column densities or the CR fluxes. In the latter case, the gas templates can be used to model the gamma-ray excess under the assumption that the CR enhancement is approximately uniform within the region of the templates themselves.    

   \begin{figure}
   \centering
  \includegraphics[width=\hsize]{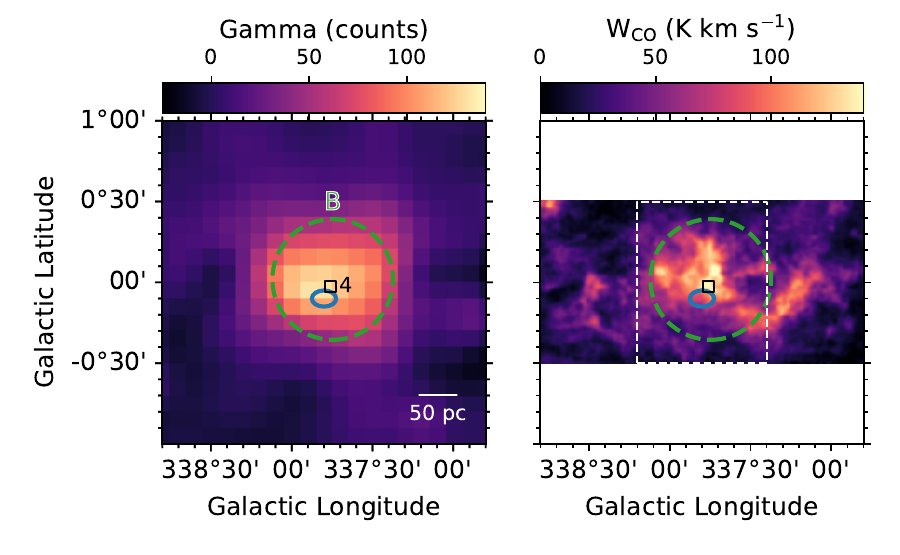}
   \caption{Region around \rightblob and SNR \object{Kes 41}. Left: photon counts that the best fit in \pbrv{\sect}~\ref{sec:geomodel} has attributed to \rightblob (dashed green circle) and the source 4FGL~J1638.5$-$4657c associated with \object{Kes 41} (shown as a square and numbered as source 4 in \pbrv{\fig}~\ref{fig:mapunassoc}). \pbrv{The counts are obtained by subtracting the model counts for all other model components from the measured counts.} Right: W$_\mathrm{CO}$  from MOPRA data in the kinematic range from  $V_\mathrm{LSR} = -70$~km~s$^{-1}$ to $-40$~km~s$^{-1}$. The blue ellipse corresponds to the radio shell of \object{Kes 41} \citep{greenUpdatedCatalogue3102025}. In the left panel we show the angular size equivalent to 50 pc at the 12~kpc distance of \object{Kes 41}. In the right plot the dashed white box delimitates the template fit to the gamma-ray data. Maps are smoothed with a 0.1\degr\ Gaussian kernel for display.
   } 
              \label{fig:kes41_maps}%
    \end{figure}

We consider first the region of \rightblob, where there is a clear overlap with CO emission in the kinematic range from  $V_\mathrm{LSR} = -70$~km~s$^{-1}$ to $-40$~km~s$^{-1}$. We produce a CO template using the high-resolution $^{12}$CO 2.6 mm line data from the MOPRA survey \citep{burtonMopraSouthernGalactic2013}. We check that across this region W$_\mathrm{CO}$ intensities estimated from the MOPRA data are consistent with those of the CO data used in the background model from CfA \citep{dameMilkyWayMolecular2001}. We restrict the template to a longitude range of 0.8\degr\ around \rightblob chosen based on the coherence of the structures in the CO $l-b-V$ data cube. According to \citet{supanNatalMolecularCloud2018} most of the gas in this velocity range within the region considered is at the distance of \object{Kes 41}, that is, $\sim$12~kpc. 

Figure~\ref{fig:kes41_maps} compares counts attributed by the fit in \pbrv{\sect}~\ref{sec:geomodel} to \rightblob and  4FGL~J1638.5$-$4657c (associated with the SNR radio shell) with the W$_\mathrm{CO}$ template. We note a good overlap of \rightblob with a high CO intensity region, and no particular enhancement of the gamma-ray emission towards the shell of the SNR or source 4FGL~J1638.5$-$4657c.

\begin{table}
\caption{Summary of the comparison between morphological models for the region of \rightblob based on CO templates. Values are relative to the model including \rightblob modeled by a disk and the point source 4FGL~J1638.5$-$4657c (source 4 in \pbrv{\fig}~\ref{fig:mapunassoc}).}             
\label{tab:ism_kes41}      
\centering                          
\begin{tabular}{l c c c}        
\hline\hline     
Model & $\Delta \ln \mathcal{L}$ & $\Delta$ d.o.f. \\
\hline
Disk + point source & 0 & 0 \\ 
W$_\mathrm{CO}$ + point source & 1.1 & $-3$ \\
W$_\mathrm{CO}$ & $-4.2$ & $-6$ \\
W$_\mathrm{CO}$ + ellipse & $-0.2$ & $-3$ \\
\hline
\end{tabular}
\end{table}
   
We fit this W$_\mathrm{CO}$ template to the gamma-ray data. For this part of the analysis all the normalization and spectral parameters of 4FGL~J1638.5$-$4657c are free whenever the source is included in the model. We first test replacing in our model the disk for \rightblob with the W$_\mathrm{CO}$ template. The results are provided in \pbrv{Table}~\ref{tab:ism_kes41}. The modest likelihood improvement, $\Delta \ln \mathcal{L}=1.1$, signifies that the disk and CO template provide equally good fits to the data. We also test replacing both the disk and source 4FGL~J1638.5$-$4657c with the W$_\mathrm{CO}$ template. This results in a likelihood decrease, $\Delta \ln \mathcal{L}=-4.1$ compared to the model with the disk and 4FGL~J1638.5$-$4657c. We also test a model which combines the W$_\mathrm{CO}$ template with an ellipse corresponding to the radio shell of the SNR: this yields a slight likelihood decrease $\Delta \ln \mathcal{L}=-0.2$ compared to the model with the disk and 4FGL~J1638.5$-$4657c. We select as baseline for the rest of this section the model with the W$_\mathrm{CO}$ template only as it has fewer degrees of freedom fit to the data and produces a likelihood close to maximum and good residuals, and also because, when including the  W$_\mathrm{CO}$ template in the fit, the significance of a source coincident or near the radio SNR shell is low, $\mathrm{TS} = 10.6$ for 4FGL~J1638.5$-$4657c or $\mathrm{TS} = 8.0$ for the radio ellipse, both $<3\sigma$.

   \begin{figure*}
   \centering
  \includegraphics[width=\hsize]{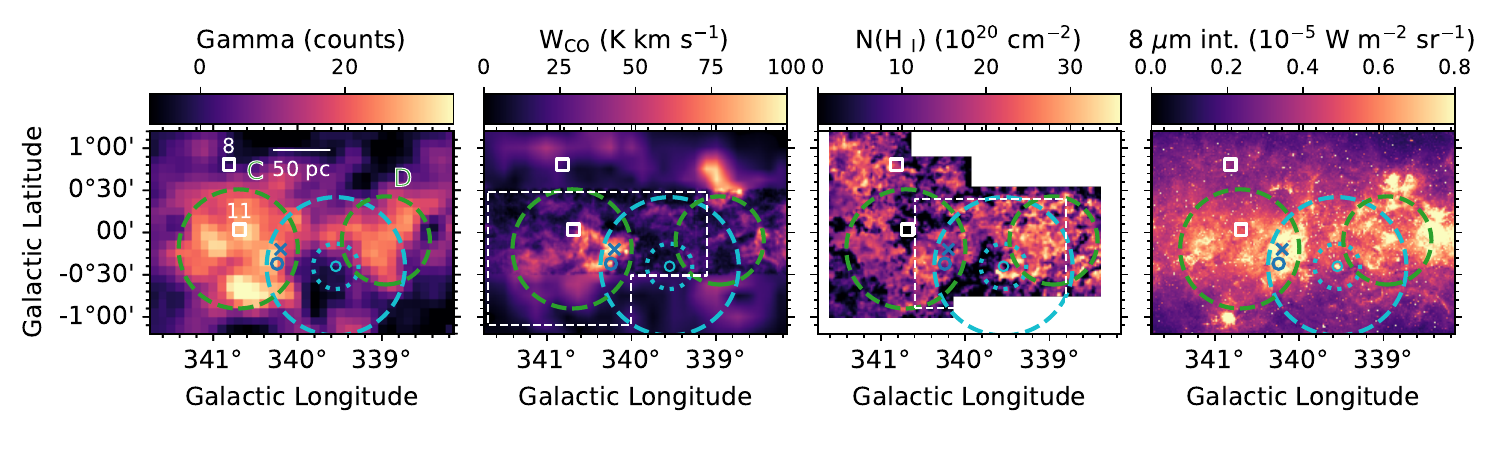}
   \caption{Region around \leftblob and \centerblob. From left to right: photon counts that the likelihood fit has attributed to  \leftblob and \centerblob (dashed green circles) based on the best-fit disk models \pbrv{in \sect~\ref{sec:geomodel} (the counts are obtained by subtracting the model counts for all other model components from the measured counts)}; CO intensity W$_\mathrm{CO}$  from MOPRA and CfA data in the kinematic range from  $V_\mathrm{LSR} = -70$~km~s$^{-1}$ to $-40$~km~s$^{-1}$; column density of atomic hydrogen, N(\hi), in the optically thin limit for lines peaking at velocities consistent with the bubble-like feature B3 \citep{kothesDistanceNeutralEnvironment2007}, highlighted by the dashed white box; 8~$\upmu$m intensity from MSX data. \latrv{Cyan circles are related to \object{Westerlund 1}: the solid circle shows the size of the stellar cluster \citep{negueruelaWesterlund1Light2022}, the dotted and dashed circles show the predicted sizes of the cluster wind termination shock and superbubble, respectively, using the values of 20~pc and 60~pc from \citet{harerUnderstandingTeVGray2023}.} The blue cross marks the position of the \hii region G340.2$-$0.2 \citep{russeilStarformingComplexesSpiral2003}. The blue circle corresponds to the cluster of two young stellar objects G340.242$-$00.370 \citep{urquhartRMSSurveyGalactic2014}. White squares mark the positions of the unassociated gamma-ray sources 4FGL~J1646.5$-$4406 and 4FGL~J1649.3$-$4441 (labeled as 11 and 8, respectively, as in \pbrv{\fig}~\ref{fig:mapunassoc}) which are kept in the model (and whose emission is subtracted from the left panel). In the left panel we show the angular size equivalent to 50 pc at the 4.23~kpc distance of \object{Westerlund 1}. In the W$_\mathrm{CO}$ panel the dashed white box delimitates the template fit to the gamma-ray data. Maps are smoothed with a 0.1\degr\ Gaussian kernel for display.
   } 
              \label{fig:shell_maps}%
    \end{figure*}

\begin{table}
\caption{Summary of the results on morphology for the region of \leftblob and \centerblob based on gas templates. Values are relative to the model M0 including \leftblob and \centerblob modeled by two disks. See the text in \pbrv{\sect}~\ref{sec:ismmodel} for details about the models.}             
\label{tab:ism_b3}      
\centering                          
\begin{tabular}{l c c c}        
\hline\hline     
Model & $\Delta \ln \mathcal{L}$ & $\Delta$ d.o.f. \\
\hline
M0 & 0 & 0 \\
M1 & $-46.9$ & $-9$ \\
M2 & $-26.7$ & $-6$ \\
M3 & $-2.4$ & $-6$ \\
\hline
\end{tabular}
\end{table}

We then move to the region of \leftblob and \centerblob. Also in this case to test quantitatively the correlation between gamma-ray emission and gas structures we fit gas templates to the gamma-ray data while removing the disks representing \leftblob and \centerblob. During this part of the analysis all spectral par\latrv{a}meters are free for 4FGL~J1649.3$-$4441 (source 11 in \pbrv{\fig}~\ref{fig:mapunassoc}, situated within \leftblob) and the normalisations are also free for the two nearby unassociated sources 4FGL~J1646.5$-$4406 and 4FGL~J1657.7$-$4520 (sources 8 and 19 in \pbrv{\fig}~\ref{fig:mapunassoc}, respectively). The results are summarized in \pbrv{Table}~\ref{tab:ism_b3}, taking the model with the two disks for \leftblob and \centerblob (M0) as reference.

Due to the clear overlap with \leftblob, we first consider a model \latrv{M1} where the two disks are replaced by a W$_\mathrm{CO}$ template in the kinematic range from $V_\mathrm{LSR} = -70$~km~s$^{-1}$ to $-40$~km~s$^{-1}$. Since \leftblob extends to higher latitudes, we merge $^{12}$CO MOPRA data with CfA data when the former are not available. We restrict the template to a region chosen based on the coherence of the structures in the CO $l-b-V$ cube around \leftblob and \centerblob (see \pbrv{\fig}~~\ref{fig:shell_maps}). A fit including only this template yields a large decrease in log-likelihood $\Delta \ln \mathcal{L} = -46.9$ but with residuals present mostly in the region of \centerblob.

To address the residual emission in the region of \centerblob, we test a model M2 in which we have two W$_\mathrm{CO}$ templates: the one of M1 plus a second template for the kinematic range from $V_\mathrm{LSR} = -130$~km~s$^{-1}$ to $-100$~km~s$^{-1}$  in the region around \centerblob. The second template is built using the same data and methodology as the first one. M2 has $\Delta \ln \mathcal{L} = -26.7$ with respect to the model with the two disks\latrv{, M0}.

We also remark that the entire region of \leftblob and \centerblob overlaps spatially with the shell of a bubble-like feature in \hi dubbed B3 in  \citet{kothesDistanceNeutralEnvironment2007}\latrv{, shown in \pbrv{\fig}~\ref{fig:shell_maps}}, which is particularly interesting to consider due to its suspected physical association with the known particle accelerator \object{Westerlund 1}. We build a column density map of \hi in the B3 shell region using data from the Southern Galactic Plane Survey \citep[SGPS,][]{mcclure-griffithsSouthernGalacticPlane2005} following the procedure described in \pbrv{Appendix}~\ref{app:hishell}. We check that across this region column densities estimated from the SGPS are consistent with those of the \hi data used in the background model. As shown in \pbrv{\fig}~\ref{fig:shell_maps}, B3 is open to the South and on the North it consists of two large emission regions, of which one overlaps with \leftblob and the peak in CO emission, the other is coincident with \centerblob.  

We test a model \latrv{M3} in which we combine the W$_\mathrm{CO}$ template in the kinematic range from $V_\mathrm{LSR} = -70$~km~s$^{-1}$ to $-40$~km~s$^{-1}$ (as in M1) and the N(\hi) template for B3. This model provides a modest decrease in log-likelihood $\Delta \ln \mathcal{L} = -2.4$ with respect to the model with two disks, with 6 fewer spatial degrees of freedom directly fit to the data. We can conclude that the model with two disks and M3 provide comparably good fits to the data.  Therefore, M3 will be taken as our reference gas template model for the region of \leftblob and \centerblob. We also note that the two unassociated sources overlapping with the gas templates remain highly significant in this model, $\mathrm{TS} = 110.4$ for 4FGL~J1649.3$-$4441 (source 11 in \pbrv{\fig}~\ref{fig:mapunassoc}) and $\mathrm{TS} = 76.9$ for 4FGL~J1646.5$-$4406 (source 8 in \pbrv{\fig}~\ref{fig:mapunassoc}).

\begin{table*}
    \caption{Spectral description of the extended sources studied in \pbrv{\sect}~\ref{sec:ismmodel}. See \pbrv{Table}~\ref{tab:geo_models_results_spec} for a description of the the columns. For all quantities the first uncertainties are statistical, while the second are systematics from the variations of the LAT effective area and interstellar background model  discussed in \pbrv{Appendix}~\ref{app:sysunc}.}             
\label{tab:gas_models_results_spec}      
\centering                          
\begin{tabular}{l c c c  c c c}        
\hline\hline     
Source  & $K$ (cm$^{-2}$ s$^{-1}$ MeV$^{-1}$) & $\alpha$ & $\beta$ & $E_\mathrm{peak}$ (GeV) & TS$_\mathrm{curv}$\\
\hline
CO \rightblob & $(2.46\pm0.11^{+0.47}_{-0.40}) \times 10^{-11}$ & $2.18\pm0.09 ^{\latrv{+0.13}}_{-0.10}$ & $0.13\pm0.04^{+0.5}_{\latrv{-0.04}}$ & $0.5\pm0.2^{+0.0}_{-0.4}$ & 15.0 \\
CO Source C\&D & $(9\pm2^{+6}_{-1}) \times 10^{-12}$ & $1.4\pm0.4^{\latrv{+0.7}}_{-0.0}$ & $0.6\pm0.2^{+0.2}_{\latrv{-0.4}}$ & $1.7\pm0.5^{+0.0}_{\latrv{-0.8}}$ & 29.4\\
\hi B3 shell & $(2.0\pm0.2^{+0.5}_{-1.0}) \times 10^{-11}$ & $1.8\pm0.4 ^{+0.5}_{-0.0}$ & $0.8\pm0.3 ^{+1.0}_{-0.1}$ & $1.2\pm0.2^{+0.0}_{-0.3}$ & 12.8\\
\hline
\end{tabular}
\end{table*}

We verify that the emission associated with the gas templates shows significant spectral curvature (TS$_\mathrm{curv} > 9$). The spectral parameters of the emission attributed by the fit to the gas templates are provided in \pbrv{Table}~\ref{tab:gas_models_results_spec}. 

\subsection{PS maps}

   \begin{figure}
   \centering
  \includegraphics[width=\hsize]{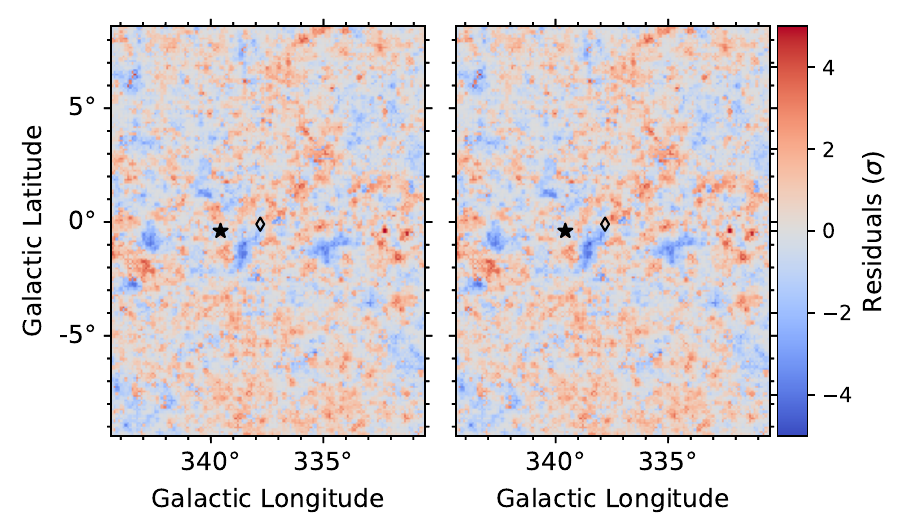}
   \caption{Data--model residuals evaluated \latrv{for the hypothesis of point-like emission} based on the method from \citet{bruelNewMethodPerform2021}. Left panel: the best model from \pbrv{\sect}~\ref{sec:geomodel} featuring extended disks. Right panel: model including gas templates from \pbrv{\sect}~\ref{sec:ismmodel}. The star marks the position of the \object{Westerlund 1} MSC \citep{tarricq3DKinematicsAge2021}, while the diamond corresponds to SNR \object{Kes 41} \citep{greenUpdatedCatalogue3102025}.
   } 
              \label{fig:psmaps}%
    \end{figure}

Figure~\ref{fig:psmaps} shows data--model residuals for both the best model from \pbrv{\sect}~\ref{sec:geomodel} with extended disks and the model in which the new extended components are modeled using gas templates in \pbrv{\sect}~~\ref{sec:ismmodel}. The distribution of the residuals amplitude is slightly larger than a normal distribution, with few excesses $> 5\sigma$ possibly indicating unmodeled emission at the rims of extended sources at the border of the ROI. \latrv{We remark the presence of spatial structures in the residuals, including} negative residuals $\lesssim -3.8\sigma$ around $l=338.6\degr$, $b = -1.4\degr$ near \object{Westerlund~1} \latrv{and residuals at angular scales $> 1\degr$ corresponding to diffuse emission of unknown origin for which the \gaussian only provides a rough description}. Overall, the quality of the residuals can be considered satisfactory for a complex region in the Galactic plane. \latrv{The large-scale structures in the residuals are one of the sources of systematic uncertainties included in assessment described in \pbrv{Appendix}~\ref{app:sysunc}.}

\subsection{Spectral energy distributions of the extended sources}

We extract spectral energy distributions (SEDs) \irii{in narrow energy bins} for the extended sources found in our analysis. \irii{For the SED points computation we keep free the normalization parameters of all the components adjusted in the broadband fit, including \pbrv{sources and} interstellar emission components.}

The \gaussian captures large-scale residuals not included in other model components and its properties vary according to the specific variation of the sky model considered. Therefore we only characterize the associated systematic uncertainties for smaller-scale components (see \pbrv{Appendix}~\ref{app:sysunc}). Conversely, a robust characterization of the \gaussian itself is beyond the scope of this article and we leave it for future publications. \pbrv{Appendix}~\ref{app:sedrout} shows that our results for the H.E.S.S. ring and \outflow are broadly consistent with those in \citet{lemoinegoumardFermiWd12025}. For the following, we focus on the newly detected extended emission components.

\begin{figure*}
\centering
\includegraphics[width=0.75\hsize]{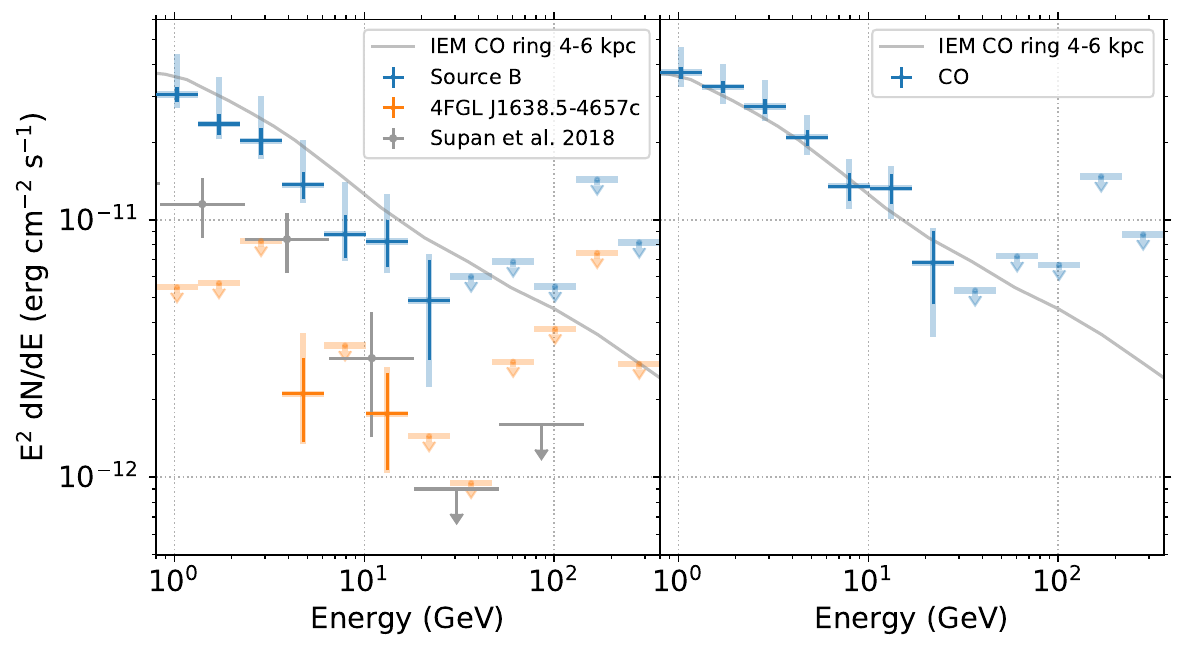}
\caption{SED of the sources in the region of \rightblob. Left panel: the pointlike source 4FGL~J1638.5$-$4657c and \rightblob in the model of \pbrv{\sect}~\ref{sec:geomodel}. Right panel: emission from the CO template in the model of \pbrv{\sect}~\ref{sec:ismmodel}. \irii{Solid error bars represent statistical uncertainties, while the semi-transparent bars show} the sum in quadrature of statistical and systematic uncertainties (see \pbrv{Appendix}~\ref{app:sysunc} for a description of systematic uncertainty sources considered). \irii{We show 95\% confidence-level upper limits} whenever the component has $\mathrm{TS}<9$ in the corresponding energy bin, or its estimated flux is consistent with zero within the $2\sigma$ statistical uncertainty. \irii{The upper limits shown are the most conservative for all model variations corresponding to different sources of systematic uncertainties.} The energy axis is truncated at 360~GeV because emission from these components is negligible at higher energies.  We also include the SED of the interstellar emission model (IEM) CO component in the ring corresponding to Galactocentric radii from 4 to 6 kpc (kinematic range of \object{Kes 41}) in the square region shown in \pbrv{\fig}~\ref{fig:kes41_maps} and, in the left panel, the SED of the pointlike gamma-ray source associated to \object{Kes 41} from \citet{supanUnidentifiedGrayEmission2018}.} 
          \label{fig:sedkes41}%
\end{figure*}
    
Figure~\ref{fig:sedkes41} shows the \latrv{the spectral energy distribution (SED)} of 4FGL~J1638.5$-$4657c, \latrv{spatially} associated with the shell of \object{Kes 41} (source 4 in \pbrv{\fig}s~\ref{fig:mapunassoc} and~\ref{fig:kes41_maps}), and of \rightblob. \irii{The large systematic error bars at low energies are mainly driven by the strong influence of the interstellar emission model on the properties of the source.}

On \irii{the} one hand, when extended emission is taken into account the flux of the pointlike source near the SNR shell is reduced by an order of magnitude at GeV energies (where the LAT PSF is comparable with the angular size of the extended emission) with respect to previous publications \citep{liuGeVGrayEmission2015,supanNatalMolecularCloud2018}. On the other hand, the total emission captured by the CO template exceeds by a factor $\sim$2 to 3 the flux attributed to the shell of \object{Kes 41} in  \citet{liuGeVGrayEmission2015,supanUnidentifiedGrayEmission2018}.

Figure~\ref{fig:sedkes41} also shows that the gamma-ray spectrum of the extended emission in the vicinity of \object{Kes 41} closely resembles below 10 GeV the one of the interstellar background attributed to the large-scale CR population. The additional extended emission on top of the background found in our analysis has a comparable intensity. Therefore, one cannot exclude that the emission is due to gas interacting with the large-scale CR population not properly taken into account in the background model. Notably, variations of a few of the CO-to-H$_2$ conversion factor are known to occur both on scales of several kpc across the Milky Way \citep[e.g.,][and references therein]{grenierNineLivesCosmic2015} and on scales of few hundreds pc for clouds with different physical properties \citep[e.g.,][]{remyCosmicRaysGas2017}.

   \begin{figure*}
   \centering
  \includegraphics[width=0.75\hsize]{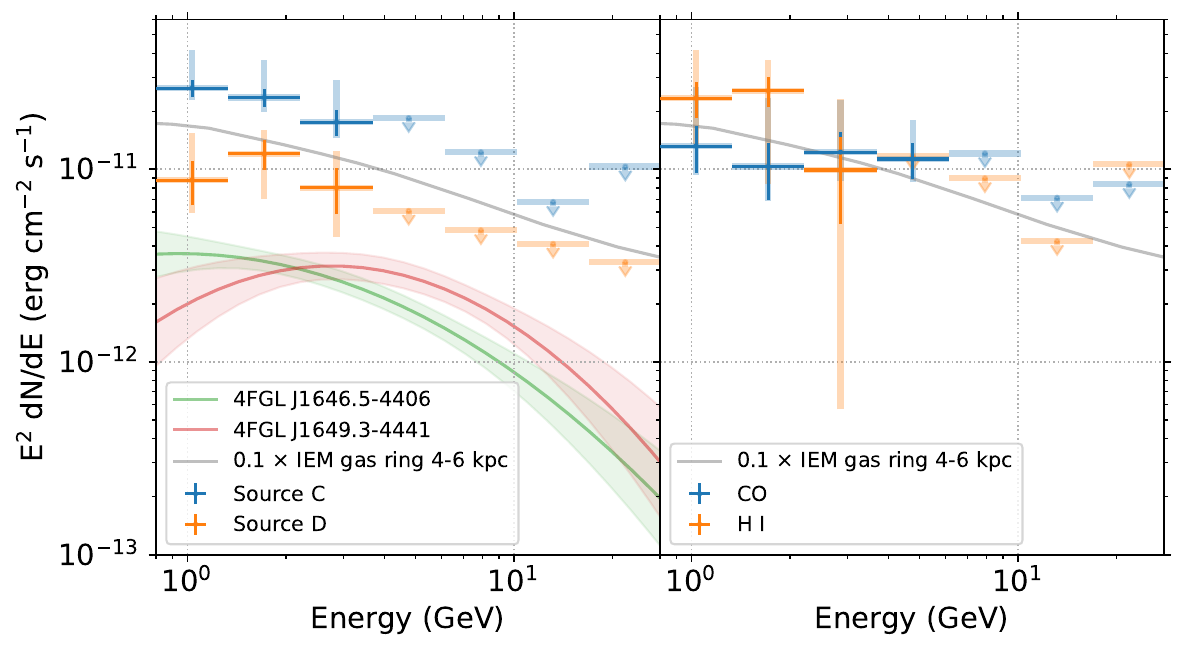}
   \caption{SEDs of the extended sources in the region of \leftblob and \centerblob. Left panel: SEDs of \leftblob and \centerblob, modeled using disks in \pbrv{\sect}~\ref{sec:geomodel}. Right panel: SEDs of the CO and \hi templates from M3 in \pbrv{\sect}~\ref{sec:ismmodel}. See \pbrv{\fig}~\ref{fig:sedkes41} for explanations about the uncertainty bars. The energy axis is truncated at 28~GeV because emission from these components is negligible at higher energies. Furthermore, we show the SED of the interstellar emission model (IEM) gas components in the ring corresponding to Galactocentric radii from 4 to 6 kpc (kinematic range of the gas templates) in a square region from $l=338.4\degr$ to $341\degr$ and $b=-1\degr$ to $0.5\degr$, scaled by 10\% for easier visual comparison. In the left panel we also show the \irii{best spectral fit over the entire energy range with its statistical uncertainties for} the two unassociated sources that overlap with \leftblob 4FGL~J1649.3$-$4441 and 4FGL~J1646.5$-$4406 (labeled as 11 and 8, respectively, in \pbrv{\fig}~\ref{fig:mapunassoc} and~\ref{fig:shell_maps}).} 
              \label{fig:sedshell}%
    \end{figure*}
    
Figure~\ref{fig:sedshell} shows the SED of the extended emission components in the region of \leftblob and \centerblob. \irii{Also in this case, the large systematic error bars at low energies are mainly driven by interstellar emission model uncertainties.} The extended emission is decomposed either in two disks (\leftblob and \centerblob, see \pbrv{\sect}~\ref{sec:geomodel}) or in two gas templates (\hi and CO, see \pbrv{\sect}~\ref{sec:ismmodel}). \jrone{The sum of the fluxes of either the two disks or the two gas templates yields a total flux} consistent within $\lesssim 15\%$ for energies below 10 GeV where the emission is significant. Additionally, in either case both components are soft (a power-law fit yields a spectral index $>2.5$). 

Figure~\ref{fig:sedshell} also compares the spectrum of the extended emission from \leftblob and \centerblob with the spectrum of the interstellar background gas components in the same kinematic range (corresponding to Galactocentric radii from 4 to 6 kpc). The extended emission detected in our analysis corresponds to $\lesssim 30\%$ of the background flux and has a similar spectral shape below 10 GeV.  The gas in this region may be subject to extreme conditions as suggested by the presence of photodissociation regions (PDRs) on these lines of sight (8~$\upmu$m emission in \pbrv{\fig}~\ref{fig:shell_maps}). Missing gas due to the presence of large amounts of ionized hydrogen, out-of-equilibrium \hi or molecular hydrogen in regions where CO is efficiently photodissociated could therefore explain the emission we detect just from interactions with the large-scale CR population in the Milky Way.

\section{Discussion}\label{sec:discussion}

We have detected three new extended sources with soft spectra in the vicinity of \object{Kes~41} and \object{Westerlund 1} overlapping with regions of high gas column densities in the Galactic plane. As discussed above, we cannot rule out that these sources are due to emission from interactions of the large-scale population of CRs with interstellar gas poorly modeled in the interstellar background model we use. Nevertheless, in this section we discuss the possible alternative hypothesis that the extended sources are due to localized CR excesses injected by the known particle accelerators in the region. 

\subsection{\rightblob}

The new extended component \rightblob is well correlated with the natal cloud of \object{Kes 41} (\pbrv{\sect}~\ref{sec:ismmodel}). Previous LAT analyses of this SNR \citep{liuGeVGrayEmission2015,supanUnidentifiedGrayEmission2018} were featuring a pointlike source near the SNR shell and overlooked the presence of several nearby unassociated gamma-ray sources overlapping the surrounding molecular cloud. Our analysis shows that extended emission associated with the cloud is statistically preferred over four nearby pointlike sources (\pbrv{\sect}~\ref{sec:geomodel}) and that a source coincident with or near the shell is not significant when the extended emission from the cloud is modeled using a CO template (\pbrv{\sect}~\ref{sec:ismmodel}).  Therefore, we conclude that there is no strong evidence for gamma-ray emission specifically associated with the $9'\times6'$ radio-emitting shell as assumed in \citet{liuGeVGrayEmission2015,supanUnidentifiedGrayEmission2018}. On the contrary, the emission appears rather to arise from the entire molecular cloud.

\latrv{W}e assume that a single population of particles interacts with the cloud, for which we assume a volume density of protons of 460~cm$^{-3}$ corresponding to the ellipse E2 in \citet{supanNatalMolecularCloud2018}. We note that this is smaller \latrv{than} the value of 950~cm$^{-3}$ used to interpret the gamma-ray spectrum by \citet{supanUnidentifiedGrayEmission2018}, who assume that the gamma-ray emission is produced in the dense gas clumps near the shell. We use \texttt{naima}\footnote{See \citet{zabalzaNaimaPythonPackage2015}. For the entire paper we will use the cross-sections from \cite{kafexhiuParametrizationGammarayProduction2014} for gamma rays from pion decay, from  \cite{baringRadioGammaRayEmission1999} for non-thermal Bremsstrahlung, and from \cite{khangulyanSimpleAnalyticalApproximations2014} for IC emission.} to estimate the properties of the underlying particle population. If we assume that gamma rays are produced by pion decay from hadronic interactions, we obtain a proton spectral index $2.3 \pm 0.2$ and a drop at an energy of $150^{+200}_{-70}$~GeV with a proton energy above 1~GeV $W_p = 4.4 \times 10^{49}$~erg.  If we assume that gamma rays are produced by electron Bremsstrahlung, we obtain an electron spectral index $2.5^{-0.3}_{+0.2}$ with a hint for a drop in the particle spectrum (Bayesian Information Criterion, BIC, smaller by 0.5) at an energy around 100~GeV. The total electron energy above 1~GeV is $W_e = 3 \times 10^{48}$~erg. These particle energies correspond to $\lesssim 5\%$ of the supernova explosion producing \object{Kes 41}, and are therefore in good agreement with expectations of diffusive shock-acceleration theory. For the standard ratio of protons to electrons in CRs the gamma-ray emission should be hadron dominated. Consistently with \citet{liuGeVGrayEmission2015,supanUnidentifiedGrayEmission2018}, we find that an inverse-Compton origin of the gamma-ray emission would require a total electron energy exceeding $10^{51}$~erg, i.e., comparable to the total energy of the \object{Kes 41} supernova. This scenario can therefore be ruled out if \object{Kes 41} is the source of the particles.

\latrv{In the evolutionary scenario favored for \object{Kes 41} by \citet{zhangMetalenrichedThermalComposite2015}, when the shock encounters the weakly ionized material outside the cavity plasma waves are damped, which causes a release of accelerated particles as discussed in  \citet{ohira2011}. Assuming representative values for a core-collapse supernova (ejecta mass of $5 M_\odot$ and initial shock velocity of 5000~km~s$^{-1}$), a cavity radius of 13~pc and a gas density in the cavity of 0.1~cm$^{-3}$ \citep{zhangMetalenrichedThermalComposite2015}, particle release should have occurred $\sim$3~kyr after the SNR birth. The radius of the cloud and gamma-ray emitting region is approximately 0.4\degr, corresponding to 80~pc at the far kinematic distance of 12~kpc. Assuming a diffusion coefficients of the same order that is inferred for large-scale propagation at GeV energies \citep[e.g.,][]{strongCosmicRayPropagation2007}, after their release the particles must have propagated over a time $> 2$~kyr to illuminate the entire cloud. This physical picture is consistent with the morphology of the gamma-ray emission which is associated with the entire molecular cloud rather than the SNR shell.}

\latrv{Following \citet{ohira2011}, the maximum energy of particles injected in the cloud is $\sim$0.5~PeV.} \latrv{The spectral break due to the finite size of the source discussed in \citet{ohira2011} falls below the energy range probed by LAT observations after a few hundred years of propagation. As shown in \pbrv{Appendix}~\ref{app:ptransport}, \pbrv{\fig}~\ref{fig:kes41transport}, the proton cooling time due to proton-proton collisions exceeds the ionization age of the system, while synchrotron cooling of electrons can hardly explain a drop in the spectrum around 100~GeV. Conversely, a break in the energy range inferred from the gamma-ray spectrum is expected after 1 to 2~kyr of propagation time due to escape} from the 80~pc cloud for a diffusion coefficient close to the average in the ISM at large.

While this suggests a possible explanation for the \latrv{morphology and spectrum of the gamma-ray emission, the scenario we discussed is based on simple considerations and requires the poorly constrained age of the system to lie within a narrow fraction of the currently allowed range. Therefore, more in-depth observational and theoretical work is needed to test its viability. } Alternatively, \citet{supanNatalMolecularCloud2018} show that the natal cloud of \object{Kes 41} is characterized by widespread massive star-formation activity which could result in distributed particle injection across the emitting region.

\subsection{\leftblob and \centerblob}

The new extended components \leftblob and \centerblob overlap with regions of high gas column densities in the Galactic plane (\pbrv{\sect}~\ref{sec:ismmodel}). In this section we speculate on the hypothesis that a localized CR excess might be related to the most powerful particle accelerator known towards these lines of sight, i.e., \object{Westerlund 1} since, as discussed in \pbrv{Appendix}~\ref{app:shell_alt}, there are no other more obvious sources that could power it. This hypothesis is energetically viable since the total flux of \leftblob and \centerblob in the model from \pbrv{\sect}~\ref{sec:geomodel} is $(7.0 \pm 0.5) \times 10^{-11}$~erg~cm$^{-2}~$s$^{-1}$ (statistical uncertainties only), which, at the distance of 4.23~kpc, represents a small fraction of order $10^{-4}$ of the MSC wind mechanical power. 

The CO and \hi templates providing the best fit to the data (M3 in \pbrv{\sect}~\ref{sec:ismmodel}, see also \pbrv{\fig}~\ref{fig:shell_maps}) correspond to interstellar gas in the same kinematic range of \object{Westerlund~1}. A physical association between the \hi B3 bubble and \object{Westerlund~1} was proposed in \citet{kothesDistanceNeutralEnvironment2007}, who suggest that B3 represents the stellar wind bubble created by the MSC.  In this scenario, the gas in the CO and \hi templates located at the edge of B3 might be (at least in part) dense material located at or near the superbubble shell. The average angular distance of \hi in the B3 shell to \object{Westerlund 1} is $\sim$0.7\degr, which corresponds to a physical size of 50~pc, slightly smaller but comparable to the expected size of the superbubble according to \citet{harerUnderstandingTeVGray2023} considering the many modeling uncertainties and projection effects. 

Intriguingly, we note the presence of bright PDRs in 8~$\upmu$m emission towards these lines of sight. We also remark the presence of the \hii region G340.2$-$0.2 \citep{russeilStarformingComplexesSpiral2003} and the cluster of two young stellar objects G340.242$-$00.370 \citep{urquhartRMSSurveyGalactic2014}, both in the same kinematic range of B3 and coincident with the peak in CO emission at the border of B3 as well as a peak in 8~$\upmu$m emission  (see \pbrv{\fig}~\ref{fig:shell_maps}). Both \irii{PDRs} and triggered star-formation are known to arise at the dense rims of bubbles around star-forming regions \citep[e.g.,][]{helouAnatomyStarFormation2004,dirienzo2012}.

The gamma-ray spectra associated with CO and \hi\ in \pbrv{\fig}~\ref{fig:sedshell} imply similar CR fluxes across the region spanned by the templates. The thickness in the plane of the sky of the high-column density region traced by the templates in the direction to \object{Westerlund~1} is $\sim$1.5\degr, equivalent to 100~pc at a distance of 4.2~kpc. In the model by \citet{harerUnderstandingTeVGray2023} the interior of the bubble is efficiently filled with particles advected from the MSC wind termination shock. For particles to diffuse through the 100-pc region near and beyond the superbubble shell over the 4~Myr lifetime of the MSC, we need a diffusion coefficient $\geq 2 \times 10^{26}$~cm$^2$~s$^{-1}$. However, we remark that there is not a perfect correspondence between the morphology of the gamma-ray emission and the gas templates (\pbrv{\fig}~\ref{fig:shell_maps}). Thus, it is not straightforward to conclude if the morphology of the emission is consistent with the scenario discussed.  

For the following discussion we assume that a single population of particles radiates in the region of \leftblob and \centerblob, for which we take a reference volume density of protons of 150~cm$^{-3}$ (\pbrv{Appendix}~\ref{app:shell_gas_dens}).
We use \texttt{naima} to estimate the properties of the particle population producing the emission from \leftblob and \centerblob. First we consider the hypothesis that gamma rays are produced by pion decay from hadronic interactions. For an unconstrained fit we infer proton spectral indices of $2.8 \pm 0.4$ and $3.1 \pm 0.6$ for \leftblob and \centerblob, respectively. These spectra are markedly steep compared with predictions from models for particles accelerated by the MSC wind termination shock. If we set a uniform prior for the spectral index in the 2-2.5 range suggested by models \jrone{that explain the H.E.S.S. data} \citep{harerUnderstandingTeVGray2023}, \jrone{comparison of models with and without a spectral drop shows that a model including a spectral drop at $\sim$50 GeV is favored, with a Bayesian Information Criterion (BIC) lower by 2.8. In this case}, the total proton energy above 1~GeV is  $W_p = 3 \times 10^{49} \;(n_\mathrm{H} /150\,\mathrm{cm}^{-3})^{-1}$~erg. If we assume that gamma rays are produced by electron Bremsstrahlung, we obtain similar conclusions. For an unconstrained fit we infer electron spectral indices of $2.9 \pm 0.5$ and $3.0 \pm 0.9$ for \leftblob and \centerblob, respectively. For a uniform prior on the spectral index in the 2–2.5 range, a model \jrone{including a spectral drop at $\sim$30 GeV is favored over the model without a spectral drop, with BIC lower by 2.7}. For a uniform prior on the spectral index in the 2-2.5 range, \irii{a model with a spectral drop at $\sim$30~GeV is favored} (BIC smaller by 2.7). For the latter case, the total electron energy above 1~GeV is  $W_e = 10^{48} \;(n_\mathrm{H} /150\,\mathrm{cm}^{-3})^{-1}$~erg. For a standard proton-to-electron ratio from diffusive-shock-acceleration theory the gamma-ray emission should therefore be hadron dominated. We note that an inverse-Compton origin of the emission cannot be excluded, but would also require a drop in the particle spectrum at energies $<100$~GeV. The region near the superbubble edge might harbor enhanced radiation fields, but even with an interstellar radiation field enhanced by a factor 10 with respect to the Galactic average, \jrone{we find that leptonic radiation is} Bremsstrahlung-dominated for gas densities of order 150~cm$^{-3}$.  

In summary, we established that, for particles accelerated at the MSC wind termination shock \jrone{and responsible for the TeV emission}, a drop in the spectrum at $\sim$50 GeV ($\sim$30 GeV) is necessary for protons (electrons) in order to reproduce the \jrone{LAT} gamma-ray spectrum \jrone{near the edge of the superbubble in the Galactic plane}, in particular to be compatible with the upper limits at energies above a few GeV. However, \jrone{H.E.S.S. data do not show evidence of such a drop} in the injected spectrum close to the wind termination shock \citep{aharonianDeepSpectromorphologicalStudy2022, harerUnderstandingTeVGray2023} nor \jrone{do the LAT data} in the region of the outflow at the edge of the Galactic disk \citep{lemoinegoumardFermiWd12025}. Possibilities to explain the drop in particle spectrum include a cooling break for electrons due to radiative losses during transport from the wind termination shock to the edge of the superbubble \citep{harerUnderstandingTeVGray2023}, or, for protons, a transition between a regime in which they lose all their energy by radiating inside the dense gas region to a regime in which they escape before cooling significantly (see \pbrv{Appendix}~\ref{app:ptransport}). An alternative physical mechanism to explain the observed spectrum may be reacceleration, which produces a characteristic ``Fermi-II bump'' at GeV energies  \citep[e.g.,][]{ferrandShapeSpectrumCosmic2010,vieuCosmicRay2022}.

In summary, there are potential physical mechanisms that may explain the properties of the gamma-ray emission from \leftblob and \centerblob as a result of particles accelerated by \object{Westerlund 1}. However, our results do not make it possible to reach firm conclusions. To this end, more detailed physical modeling and more accurate observational constraints are needed.

\section{Summary and conclusions}\label{sec:conclusions}

The paper presents an analysis of 16.4~years of \F~LAT observations at energies $>800$~MeV towards the \object{Westerlund 1} MSC and the \object{Kes 41} SNR. The 4FGL-DR4 catalog contains 18 unassociated gamma-ray sources within 2.5\degr\ from the MSC. We show that replacing 15 of them with 6 extended sources is statistically favored. Three of these extended sources with hard spectra were already discussed in our previous paper based on an analysis at energies $> 3$~GeV \citep{lemoinegoumardFermiWd12025}. This paper presents the detection of three additional extended sources with soft spectra coincident with regions of high gas column density in the Galactic plane.

\rightblob  is well associated spatially with the natal cloud of the SNR \object{Kes 41} at a distance of 12~kpc.
The detection of extended emission from the cloud renders marginal the detection of pointlike emission towards the shell of the SNR claimed from previous analyses. The extended emission can be explained by either an underestimate of the mass of the cloud by a factor of 2 in the interstellar background model or the conversion of $\lesssim 5\%$ of the supernova explosion energy to accelerated particles that have left the SNR shell and are diffusing through the cloud, producing gamma rays either via nucleon collisions or electron Bremsstrahlung. For the local injection of particles, the soft spectrum of the emission may be explained by diffusive escape of the higher-energy particles from the cloud.

Two more extended components, \leftblob and \centerblob, are equally well modeled by two disks or by two gas templates that account for atomic and molecular gas in a region within $\sim$100~pc from the edge of the \object{Westerlund 1} superbubble. The fluxes of these extended emission components, corresponding to $\lesssim 30\%$ of the interstellar background, could easily be explained by poorly modeled gas interacting with the large-scale CR population. However, also particles injected or reaccelerated in the  \object{Westerlund 1} superbubble could explain the measured energy flux equivalent to a small fraction of order $10^{-4}$ of the MSC wind mechanical power. In the latter case, the soft spectrum might be explained by different physical mechanisms: Bremsstrahlung of electrons cooled during the transport from the MSC wind termination shock to the superbubble shell; decay of pions generated in collisions of nuclei accelerated at the MSC wind termination shock and that lose all their energy in the dense gas region near the superbubble edge at low energies while they can diffusively escape at higher energies; particles with a bumpy spectrum due to second-order Fermi acceleration in the superbubble.

Our results strengthen the evidence for gamma-ray emission structures arising at intermediate spatial scales between isolated objects and the large-scale diffuse emission from the ISM. For \object{Westerlund 1} the morphology of the extended emission surrounding the MSC is particularly complex and strongly energy dependent with a TeV ring possibly tracing the active acceleration region near the MSC wind termination shock, a hard GeV component that can be interpreted as a nascent cosmic-ray loaded outflow towards the edge of the Galactic disk, and, as shown in this paper, soft GeV emission from the dense neutral gas near the edge of the superbubble.  

Our results also bear an importance to understand the nature of unassociated \F-LAT sources. Overall, 1/3 of the sources in 4FGL-DR4 remain unassociated with known multiwavelength emitters. However, the fraction of unassociated sources reaches 59\% in the Galactic plane ($|b|<10\degr$), of which 74\% have soft spectra (power-law index $> 2.4$) that set them apart from known classes of gamma-ray emitters   \citep{abdollahiFermiLargeArea2020b,LAT2026GalacticUnassociated}. We have shown that in the region studied in this paper the majority of Galactic unassociated sources can in fact be attributed to extended emission (even for those modeled as point-like in catalogs) and that links with \irii{known} objects can be drawn not looking at direct spatial association with a particle accelerator but rather considering particle transport and interstellar structures in their vicinities. Interestingly, the unassociated sources/extended emission components overlapping with the natal cloud of \object{Kes 41} and the neutral gas near the \object{Westerlund 1} superbubble have soft spectra that can be explained either by gas missing in the interstellar background model or by the physical processes driving particle transport and cooling around the injection site.

The analysis presented in this paper \latrv{provides the most comprehensive characterization to date of extended GeV gamma-ray emission in the vicinity of \object{Westerlund 1} and \object{Kes 41}, offering a solid foundation for future studies. However, it remains} limited by the angular resolution of the current \F~LAT dataset and by the interstellar background model adopted. An improved description of the interstellar background based on a separation of high-resolution gas spectra along the line of sight specifically tailored to this region and higher-resolution gamma-ray data can help to extend the analysis to lower energies and discriminate between the hypotheses presented to explain the soft extended emission around \object{Westerlund 1} and \object{Kes 41}. Higher-resolution more sensitive observations could also help to detect the high-energy tail of the ``Fermi-II bump'' and make it possible to validate or exclude this scenario. Furthermore, the physical interpretations of the observations we discussed are based on simple considerations, but detailed modeling work is needed in order to confirm if they are viable and to constrain physical parameters of the systems. Finally, extended gamma-ray emission coincident with ISM structures in the vicinities of particle accelerators may be widespread and account, at least in part, for the large number of soft unassociated sources in LAT catalogs, which strongly motivates detailed analyses of other individual regions as well as statistical studies for the entire Galactic plane.

\begin{acknowledgements}
The \textit{Fermi} LAT Collaboration acknowledges generous ongoing support
from a number of agencies and institutes that have supported both the
development and the operation of the LAT as well as scientific data analysis.
These include the National Aeronautics and Space Administration and the
Department of Energy in the United States, the Commissariat \`a l'Energie Atomique
and the Centre National de la Recherche Scientifique / Institut National de Physique
Nucl\'eaire et de Physique des Particules in France, the Agenzia Spaziale Italiana
and the Istituto Nazionale di Fisica Nucleare in Italy, the Ministry of Education,
Culture, Sports, Science and Technology (MEXT), High Energy Accelerator Research
Organization (KEK) and Japan Aerospace Exploration Agency (JAXA) in Japan, and
the K.~A.~Wallenberg Foundation, the Swedish Research Council and the
Swedish National Space Board in Sweden.

Additional support for science analysis during the operations phase is gratefully 
acknowledged from the Istituto Nazionale di Astrofisica in Italy and the Centre 
National d'\'Etudes Spatiales in France. This work performed in part under DOE 
Contract DE-AC02-76SF00515.

This work has made use of the following open-source software packages: APLPy \citep{2012ascl.soft08017R}, GAMERA\footnote{\url{https://libgamera.github.io/GAMERA/docs/documentation.html}}, matplotlib \citep{Hunter:2007}, naima \citep{zabalzaNaimaPythonPackage2015}, numpy \citep{harris2020array}, scipy \citep{2020SciPy-NMeth}. 

This research has made use of data products and databases from the following sources: the Astrophysics Data System, funded by NASA under Cooperative Agreement 80NSSC21M0056; the ATNF pulsar catalog\footnote{\url{https://www.atnf.csiro.au/research/pulsar/psrcat} \citep{manchesterAustraliaTelescopeNational2005}}; the Midcourse Space Experiment (processing of the data was funded by the Ballistic Missile Defense Organization with additional support from NASA Office of Space Science); the Mopra radio telescope (part of the Australia Telescope National Facility which is funded by the Commonwealth of Australia for operation as a National Facility managed by CSIRO and support from the Australian Research Council, UNSW, Sydney and Monash Universities); the NASA/ IPAC Infrared Science Archive (operated by the Jet Propulsion Laboratory, California Institute of Technology, under contract with NASA);  SNRCat\footnote{\url{http://snrcat.physics.umanitoba.ca/}} \citep{ferrandCensusHighenergyObservations2012}.\latrv{The authors would like to thank I. A. Grenier, P. Bruel and J. Ballet for useful suggestions about the manuscript.}

\end{acknowledgements}

%
   \bibliographystyle{aa} 
   \bibliography{Wd1Fermi25.bib} 
   
\begin{appendix}

\section{List of remarkable sources}\label{app:sourcelist}

Table~\ref{tab:unassoc} lists the sources that are discussed individually in the text with the numbering from \pbrv{\fig}~\ref{fig:mapunassoc}, their 4FGL-DR4 names, and their classification in the catalog. 

\begin{table}
\caption{4FGL-DR4 names and catalog classification for the sources numbered in \pbrv{\fig}~\ref{fig:mapunassoc}.}.             
\label{tab:unassoc}      
\centering                          
\begin{tabular}{lcc}        
\hline\hline                 
Number & Name & Class  \\    
\hline                        
1 & 4FGL J1636.9$-$4710c &  \\
2 & 4FGL J1638.1$-$4641c &  \\
3 & 4FGL J1638.4$-$4715c &  \\
4 & 4FGL J1638.5$-$4657c & spp \\
5 & 4FGL J1639.8$-$4642c &  \\
6 & 4FGL J1644.5$-$4602c & bcu \\
7 & 4FGL J1645.8$-$4533c & unk \\
8 & 4FGL J1646.5$-$4406 &  \\
9 & 4FGL J1648.4$-$4554 &  \\
10 & 4FGL J1649.2$-$4513c &  \\
11 & 4FGL J1649.3$-$4441 &  \\
12 & 4FGL J1650.9$-$4420c &  \\
13 & 4FGL J1651.4$-$4442c &  \\
14 & 4FGL J1652.2$-$4633e &  \\
15 & 4FGL J1652.2$-$4516 &  \\
16 & 4FGL J1655.5$-$4737e &  \\
17 & 4FGL J1656.1$-$4706c &  \\
18 & 4FGL J1657.7$-$4656c &  \\
19 & 4FGL J1657.7$-$4520 &  \\
20 & 4FGL J1658.3$-$4637c &  \\
\hline                                   
\end{tabular}
\end{table}

\section{Treatment of the interstellar background and preliminary model adjustments}\label{app:prelfit}

In 4FGL-DR4 the interstellar background is modelled based on the linear combination of templates describing emission associated with atomic gas traced by the 21~cm \hi line and molecular gas traced by the 2.6~mm $^{12}$CO line split in several Galactocentric rings.  Infrared tracers of dust column density were used to correct gas column densities in directions where the combination of \hi and CO templates was either under or over-estimating the data (DNMp and DNMn components). The model also includes components to account for IC emission, unresolved sources, and diffuse emission of unknown origin not captured by the templates, known as the ``patch'' component, that is iteratively determined from residuals in the LAT data\latrv{\footnote{\url{https://fermi.gsfc.nasa.gov/ssc/data/analysis/software/aux/4fgl/Galactic_Diffuse_Emission_Model_for_the_4FGL_Catalog_Analysis.pdf}}}.

\begin{table}
\caption{Interstellar emission components individually fit to the data. Components based on \hi or CO line data are split in several ranges of Galactocentric radius $R$.}             
\label{tab:iemcomp}      
\centering                          
\begin{tabular}{cc}        
\hline\hline                 
Component type & $R$ (kpc)  \\    
\hline                        
\hi & 0.6-4 \\
\hi & 4-6 \\
\hi & 6-7 \\
\hi & 7-9 \\
\hi & 9-35 \\
CO & 0.6-4 \\
CO & 4-6 \\
CO & 6-7 \\
CO & 7-9 \\
IC & - \\
DNMp & - \\
\hline                                   
\end{tabular}
\end{table}

In the catalog as well as standard LAT analyses all the components of the interstellar emission model are combined and fit to the data by adjusting a global normalisation coefficient and, possibly, a power-law index as a spectral correction\footnote{See \url{https://fermi.gsfc.nasa.gov/ssc/data/access/lat/BackgroundModels.html}. The model used in 4FGL-DR4 is \texttt{gll\_iem\_v07.fits}.}. This is problematic for our analysis because the patch component includes structured emission towards Westerlund 1 on angular scales close to those of interest. Therefore, for our work we remove the 4FGL patch component and, for a more accurate modeling of emission in our ROI, we adjust separately to the data several model components. Among the model components we select only those contributing to emission in the ROI, and some of the components are grouped together either because they contribute a limited number of photons or their morphologies are not different enough to be effectively separated in a likelihood fit in our ROI. Based on this criteria we end up with 11 interstellar components fit to the data detailed in \pbrv{Table}~\ref{tab:iemcomp}. The DNMn component (negative correction) cannot be fit individually to the data for technical reasons, but its importance in our ROI is very modest. 

   \begin{figure}
   \centering
  \includegraphics[width=\hsize]{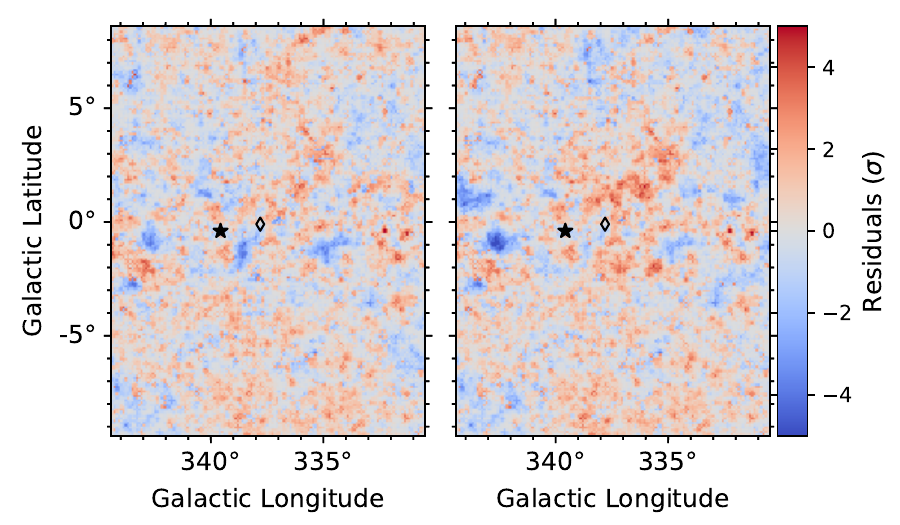}
   \caption{Data--model residuals evaluated \latrv{for the hypothesis of point-like emission} based on the method from \citet{bruelNewMethodPerform2021}. Left panel: the best model from \pbrv{\sect}~\ref{sec:geomodel} featuring the \gaussian (same as left panel of \pbrv{\fig}~\ref{fig:psmaps}). Right panel: the same model where the \gaussian is replaced by the 4FGL patch. The star marks the position of the \object{Westerlund 1} MSC \citep{tarricq3DKinematicsAge2021}, while the diamond corresponds to SNR \object{Kes 41} \citep{greenUpdatedCatalogue3102025}.
   } 
              \label{fig:psmaps_patch}%
    \end{figure}

   \begin{figure}
   \centering
   \includegraphics[width=\hsize]{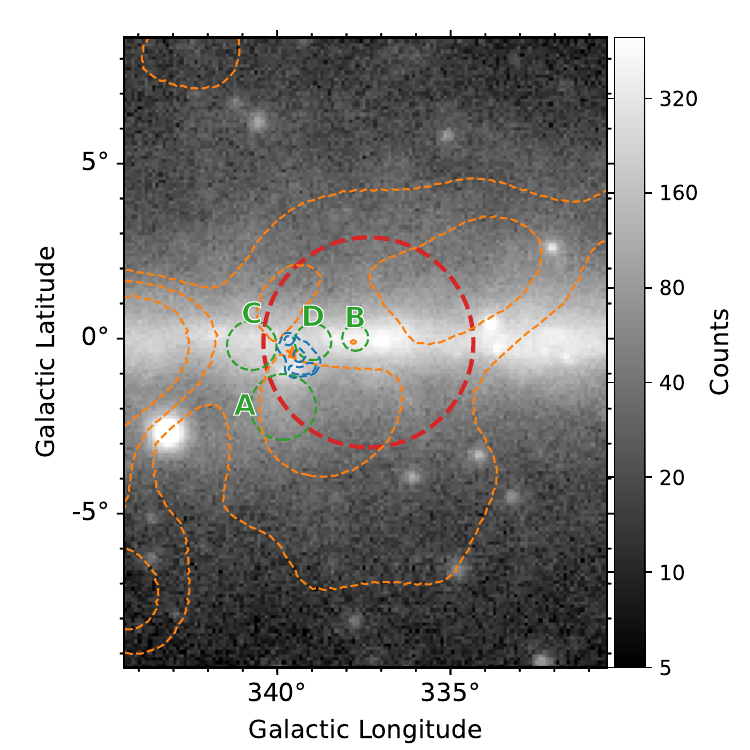}
      \caption{68\% containment  regions for extended model components studied in the paper (\pbrv{\sect}~\ref{sec:geomodel}) with the red dashed circle corresponding to the \gaussian and the orange contours corresponding to the residual-driven patch in the 4FGL-DR4 interstellar background model. See \pbrv{\fig}~\ref{fig:mapunassoc} for a description of other figure elements.
      }
      \label{fig:mappatch}
   \end{figure}

We perform a first fit to the LAT data with free normalizations for sources with $\mathrm{TS}>5000$ and interstellar background components, plus spectral parameters for the two bright sources 4FGL~J1709.7$-$4429 and 4FGL~J1620.7$-$4927. These parameters will always be let free in the subsequent analyses. This first fit reveals large-scale residuals over $\sim$10\degr\ correlated with the 4FGL patch component that was removed. Therefore we introduce an extended source to account for the large-scale residuals. We test both a disk and a 2D Gaussian morphology. The 2D Gaussian provides the best fit with $\mathrm{TS} = 808$ for 5 additional degrees of freedom (significance 19.5$\sigma$). This source replaces the patch in the 4FGL-DR4 interstellar background model and results in improved residuals (\pbrv{\fig}~\ref{fig:psmaps_patch}). Replacing the 4FGL patch with the 2D Gaussian ensures that this component is smooth on the scale of the ROI and makes it possible to iteratively adjust it along with other extended sources. We will refer to this source as \gaussian. Its position and extension for the model in \pbrv{\sect}~\ref{sec:geomodel} are shown in \pbrv{\fig}~\ref{fig:mappatch}.

At this point we note significant residuals near five extended sources with multiwavelength counterparts: 4FGL~J1615.3$-$5146e, 4FGL~J1616.2$-$5054e, 4FGL~J1633.0$-$4746e, 4FGL~J1631.6$-$4756e, and 4FGL~J1636.3$-$4731e. The morphological models for those sources do not come from multiwavelength data, they were fit to the LAT data in previous analyses based on different datasets and energy selections. Therefore, we refit all their morphological and spectral parameters. This yields $\mathrm{TS}=249.4$ for 28 additional degrees of freedom (significance 7.5$\sigma$).

Based on the results in \citet{lemoinegoumardFermiWd12025}, first we replace the three sources 4FGL~J1645.8$-$4533c \latrv{(number 7 in \pbrv{\fig}~\ref{fig:mapunassoc}, with a multiwavelength counterpart of unknown nature)}, 4FGL~J1648.4$-$4554 \latrv{(number 9, unassociated)} and 4FGL~J1644.5$-$4602c (\latrv{number 6, associated with a blazar of unknown type)} with two templates based on the H.E.S.S. flux map from \citet{aharonianDeepSpectromorphologicalStudy2022} modeling the ring-like feature around \object{Westerlund 1}. This yields $\mathrm{AIC} = -43.8$, therefore favoring the model with the extended H.E.S.S. templates. The H.E.S.S. template is split in two because the brigth\latrv{n}ess appears non uniform in the LAT energy band (see \citealt{lemoinegoumardFermiWd12025} for details). Then we replace the five unassociated sources 4FGL~J1652.2$-$4633e, 4FGL J1655.5$-$4737e, 4FGL J1656.1$-$4706c, 4FGL J1657.7$-$4656c, and 4FGL J1658.3$-$4637c  (numbered as 14, 16, 17, 18 and 20 in \pbrv{\fig}~\ref{fig:mapunassoc}) with a single extended source with a 2D Gaussian morphology whose parameters are fit to the LAT data. This yields $\mathrm{AIC} = -136.4$, therefore favoring again the model proposed in  \citet{lemoinegoumardFermiWd12025}. This extended source will hereafter be referred to as \outflow and it corresponds to the outflow from \object{Westerlund 1} discussed  in  \citet{lemoinegoumardFermiWd12025}.

\section{Evaluation of systematic uncertainties}\label{app:sysunc}

We provide an assessment of the systematic uncertainties on source morphological parameters due to the interstellar background. For our limited assessment we will consider three \latrv{families of} model variations which stand out as particularly prominent.
\begin{itemize}
\item In our analysis we individually adjust the normalization of 11 interstellar background components. Deviations from the nominal values are mostly within $\pm 25\%$, but can reach $80\%$ for a few components. Thus, we consider a model variation in which all the components have a normalization fixed to 1. 
\item In our analysis we introduce a \gaussian to replace the 4FGL patch component. The patch, however modeled, is bright across the ROI and its nature remains not understood.  Thus, we consider \latrv{two model variations. In the first one} we reintroduce the original 4FGL patch template in lieu of the \gaussian. \latrv{In the second variation, we create a customized patch obtained from the residuals from the analysis of \pbrv{\sect}~\ref{sec:geomodel} filtered to keep only structures at angular scales $> 1\degr$.}
\item We remark that the DNMp component has \jrone{hot spots coincident with \rightblob and \leftblob. Since the procedure based on dust tracer residuals used to build the DNMp component\footnote{\url{https://fermi.gsfc.nasa.gov/ssc/data/analysis/software/aux/4fgl/Galactic_Diffuse_Emission_Model_for_the_4FGL_Catalog_Analysis.pdf}} suffers from severe approximations in this complex star-forming region, we consider a model variation in which the hot spots in the DNMp component} are masked.
\end{itemize}
We estimate the systematic uncertainties as the maximum difference with respect to the baseline value obtained with these three \latrv{families of} model variations. 

We stress that the errors evaluated through this procedure do not capture all the possible sources of uncertainties, including unknown levels of \hi and CO optical thickness, CO dissociation and CO excitation gradients in the vicinity of bright MSCs, as well as large dust temperature and emissivity gradients in such environments.

For spectral properties of the gamma-ray sources, in addition to the interstellar background systematics, we also take into account uncertainties in the LAT instrument effective area. They are assessed using the method of the bracketing IRFs\footnote{We use the \texttt{min} and \texttt{max} scalings from \url{https://fermi.gsfc.nasa.gov/ssc/data/analysis/scitools/Aeff_Systematics.html}}.

Finally, we also consider uncertainties in the modeling of extended emission components related to ISM structures, assessed by comparing results obtained with the Gaussian/disk models from \pbrv{\sect}~\ref{sec:geomodel} and the gas templates from \pbrv{\sect}~\ref{sec:ismmodel}. 

Uncertainties from these three sources -- interstellar background, effective area, and source model -- are considered as statistically independent and thus summed in quadrature whenever needed.

\section{Gamma-ray maps in energy bands}\label{app:maps_ebands}

   \begin{figure*}
   \centering
  \includegraphics[width=\textwidth]{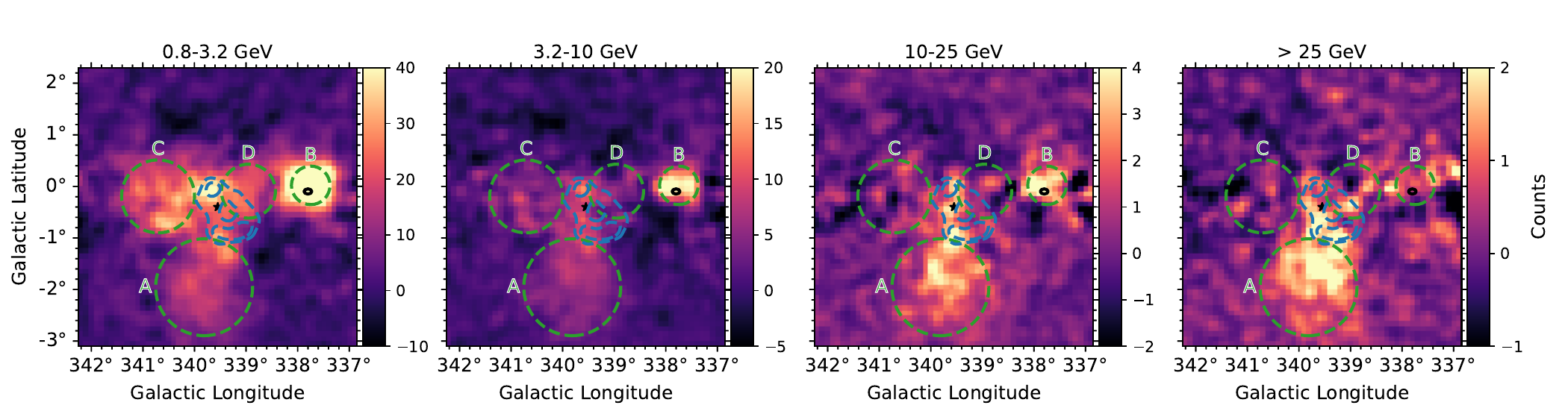}
   \caption{Photon counts that the likelihood fit has  attributed to extended sources in the vicinity of \object{Westerlund 1} and \object{Kes 41}, the two H.E.S.S. templates and sources A to D. \pbrv{The counts are obtained by subtracting the model counts for all other model components from the measured counts, thereby isolating the contribution of these objects or sources of unclear nature}. Each panel corresponds to a different energy range. Maps are smoothed with a kernel of 0.1\degr\ for display. The black star marks the position of \object{Westerlund 1} \citep{tarricq3DKinematicsAge2021}, while the black ellipse corresponds to the radio shell of SNR G337.8$-$0.1, alias \object{Kes 41} \citep{greenUpdatedCatalogue3102025}. \irii{Blue dashed contours correspond to TeV gamma-ray emission measured with H.E.S.S. towards \object{Westerlund 1} \citep{aharonianDeepSpectromorphologicalStudy2022}.}}
              \label{fig:excess_ebands}%
    \end{figure*}
    
Figure~\ref{fig:excess_ebands} shows for several energy bands the photon counts that the likelihood fit has attributed to the extended emission components, i.e., measured counts minus the best-fit model except for the extended emission components (H.E.S.S. templates, \outflow, \rightblob, \leftblob, \centerblob). As found in previous studies the H.E.S.S. ring-like feature and \outflow have hard spectra and become dominant above 25 GeV. \rightblob has a curved spectrum with emission dropping before 25 GeV. \leftblob and \centerblob have soft spectra with emission dropping below 10 GeV and which becomes most evident below 3.2~GeV.

\section{CO maps in velocity bands}\label{app:covbands}

   \begin{figure*}[p]
   \centering
  \includegraphics[height=0.9\vsize]{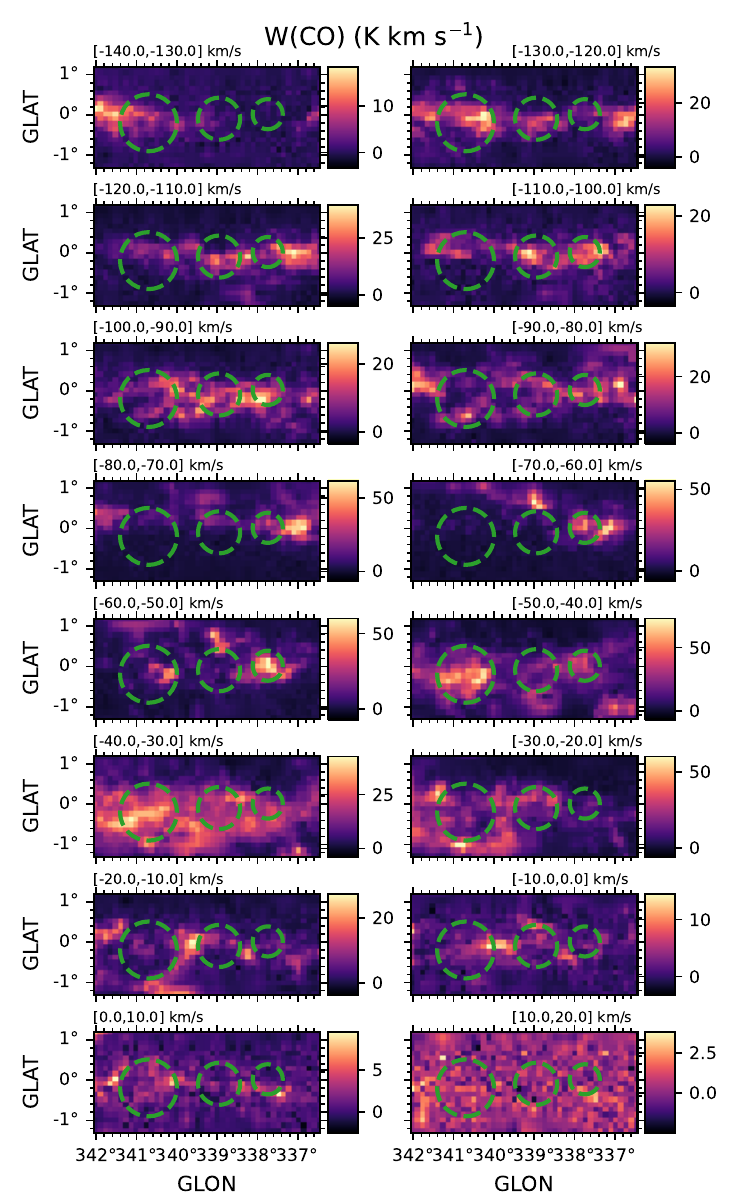}
   \caption{Velocity-integrated brightness temperature of the $^{12}$CO 2.6 mm line (W$_\mathrm{CO}$) from \citet{dameMilkyWayMolecular2001} in several velocity ranges. The circles overlaid to the CO maps correspond, from right to left, to the 68\% containment radius of \rightblob, \centerblob and \leftblob (see \pbrv{\fig}~\ref{fig:mapunassoc}).} 
              \label{fig:cofinderchart}%
    \end{figure*}

Figure~\ref{fig:cofinderchart} shows the distribution of molecular gas in the region towards \rightblob, \leftblob and \centerblob in several velocity intervals.

\section{\hi map of the B3 feature}\label{app:hishell}
Using high-resolution \hi data from the SGPS \citep{mcclure-griffithsSouthernGalacticPlane2005}, \citet{kothesDistanceNeutralEnvironment2007} report the detection of a degree-size expanding bubble dubbed B3 in the vicinity of \object{Westerlund 1} at a velocity $V_\mathrm{LSR} \sim -51$~km~s$^{-1}$. Determining a column-density template for B3 is challenging due to the superposition of multiple structures along the line of sight with a strong overlap in velocity space owing to the width of the \hi lines.

After identifying a coherent structure in $l-b-V$ space from a visual inspection of the 3D \hi cube corresponding to the bubble, we perform a spectral decomposition for the relevant lines of sight. We model the \hi spectra as a combination of pseudo-Voigt profiles similarly to \citet{remyCosmicRaysGas2017} and assign to the B3 template the entire column density for lines with a peak velocity in the range from $V_\mathrm{LSR} = -56$~km~s$^{-1}$ to $-46$~km~s$^{-1}$, chosen to isolate the coherent feature previously identified in the 3D \hi cube.

We note that the template thus obtained extends beyond the bubble B3 as identified in \citet{kothesDistanceNeutralEnvironment2007}, including emission around $l = 341\degr$ which appears connected in $l-b-V$ space with the bubble shell.

\section{Spectrum of the ring-like H.E.S.S. feature and \outflow}\label{app:sedrout}

   \begin{figure}
   \centering
  \includegraphics[width=1\hsize]{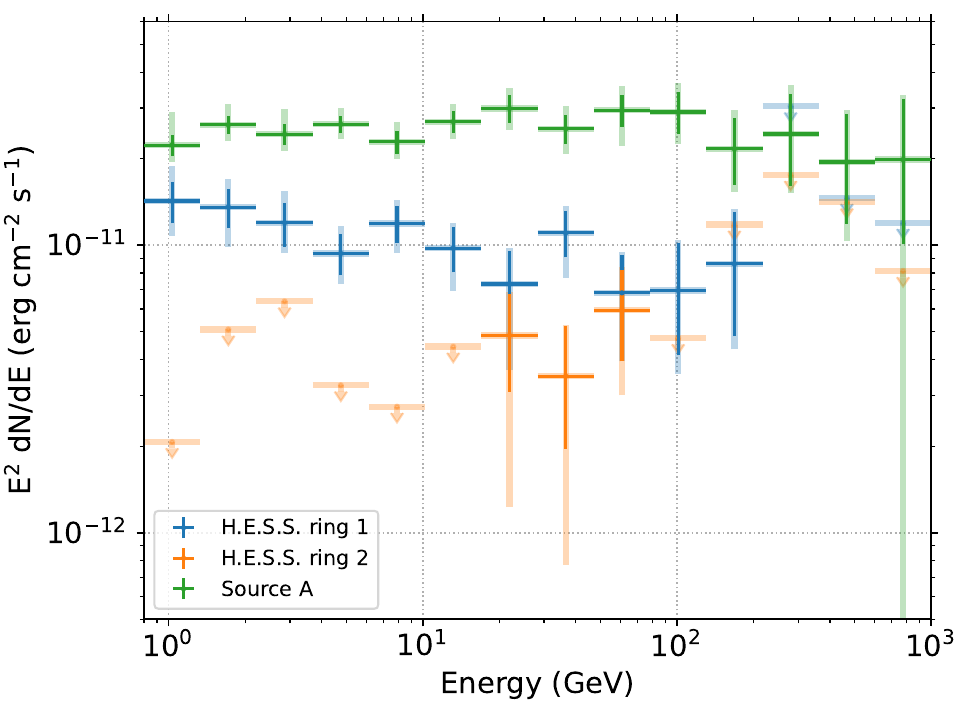}
   \caption{Spectral energy distribution (SED) of the sources associated with the H.E.S.S. ring and \outflow. See \pbrv{\fig}~\ref{fig:sedkes41} for explanations about the uncertainty bars. Reference values are from the model described in \pbrv{\sect}~\ref{sec:geomodel}, while the model from \pbrv{\sect}~\ref{sec:ismmodel} is considered to estimate the systematic uncertainties.} 
              \label{fig:sedrout}%
    \end{figure}
    
Figure~\ref{fig:sedrout} shows the SED of the components associated with the TeV ring and the outflow in the southern direction of \object{Westerlund 1} (\outflow). Overall, our results are very consistent with the previous analysis $>3$~GeV in \citet{lemoinegoumardFermiWd12025}, especially for \outflow and region~1 of the H.E.S.S. ring. Emission from region~2 of the H.E.S.S. ring is less significant in our analysis compared to \citet{lemoinegoumardFermiWd12025}, probably due to the addition of \centerblob which largely overlaps with this part of the ring. However, we note that the spectrum of \centerblob is soft, while we still find hard emission towards region~2, connecting  with the emission detected at higher energies by H.E.S.S.  

\section{Particle transport and energy \latrv{loss} timescales}\label{app:ptransport}

For the cases of interest particle transport is dominated by diffusion, that we parametrize via a diffusion coefficient $D$ as a function of particle energy $E$:
\begin{equation}
D (E) = D_0 \left(\frac{E}{1\;\mathrm{GeV}}\right)^\delta.
\end{equation}
All across this appendix the characteristic timescales for energy losses of particles are calculated using \texttt{GAMERA}\footnote{\url{http://libgamera.github.io/GAMERA/docs/main_page.html}}. In addition to the cross-section models referenced in \irii{\pbrv{\sect}}~\ref{sec:discussion} we use the formulas for synchrotron radiation given in \citet{ghiselliniSynchrotronBoiler1988}.

   \begin{figure}
   \centering
  \includegraphics[width=\hsize]{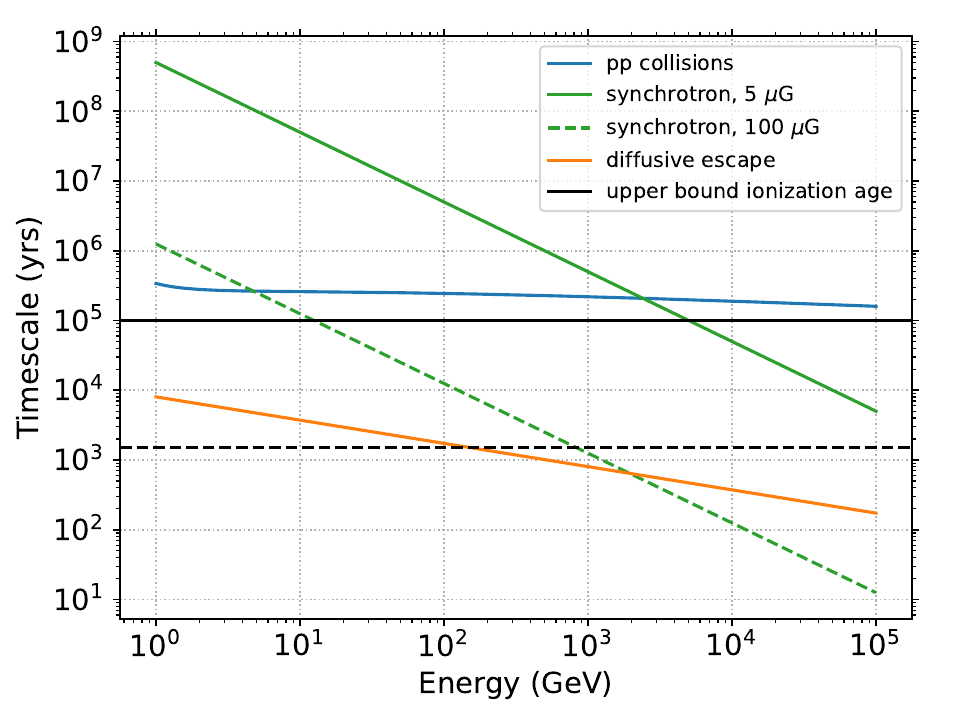}
   \caption{Timescales for particle cooling and diffusive escape in the natal cloud of \object{Kes 41} as a function of energy. The diffusive escape time is calculated for a cloud size of 80~pc and a diffusion coefficient with energy slope $\delta=1/3$ and normalization $D_0 = 6\times 10^{28}$~cm$^{2}$~s$^{-1}$. The proton cooling times are calculated for a proton density of 460~cm$^{-3}$. The electron synchrotron cooling times are calculated for strengths of the magnetic field of 5 and 100~$\upmu$G. \latrv{The solid horizontal line shows the upper bound on the ionization age of the SNR from \citet{zhangMetalenrichedThermalComposite2015}. The dashed} horizontal line shows for illustration a \latrv{propagation} time of 1.5~kyr\latrv{, for which the spectral break due to diffusive escape occurs at $\sim$150~GeV.}
   } \label{fig:kes41transport}%
    \end{figure}

In the case of the natal cloud of \object{Kes 41}, \pbrv{\fig}~\ref{fig:kes41transport} compares the timescales for diffusive escape and particle cooling. \latrv{The proton cooling timescale is larger than the upper bound on the ionization age of the SNR from \citet{zhangMetalenrichedThermalComposite2015}.} Electron cooling in a few $\upmu$G magnetic field, typical of interstellar clouds, cannot explain a cutoff at $\sim$100 GeV. \latrv{Even assuming a higher field strength of 100~$\upmu$G, a cutoff at $\sim$100 GeV requires a propagation time $>10$~kyr. On \jrone{these} timescales, for an average diffusion coefficient, a spectral break due to escape prevails. Indeed,} the figure illustrates that for a diffusion coefficient close to the interstellar average with a normalization $D_0 = 6\times 10^{28}$~cm$^{2}$~s$^{-1}$ \latrv{and a slope $\delta=1/3$}, escape from the cloud after 1.5~kyr of injection by the SNR could suppress particle fluxes $\gtrsim 150$~GeV.

   \begin{figure}
   \centering
  \includegraphics[width=\hsize]{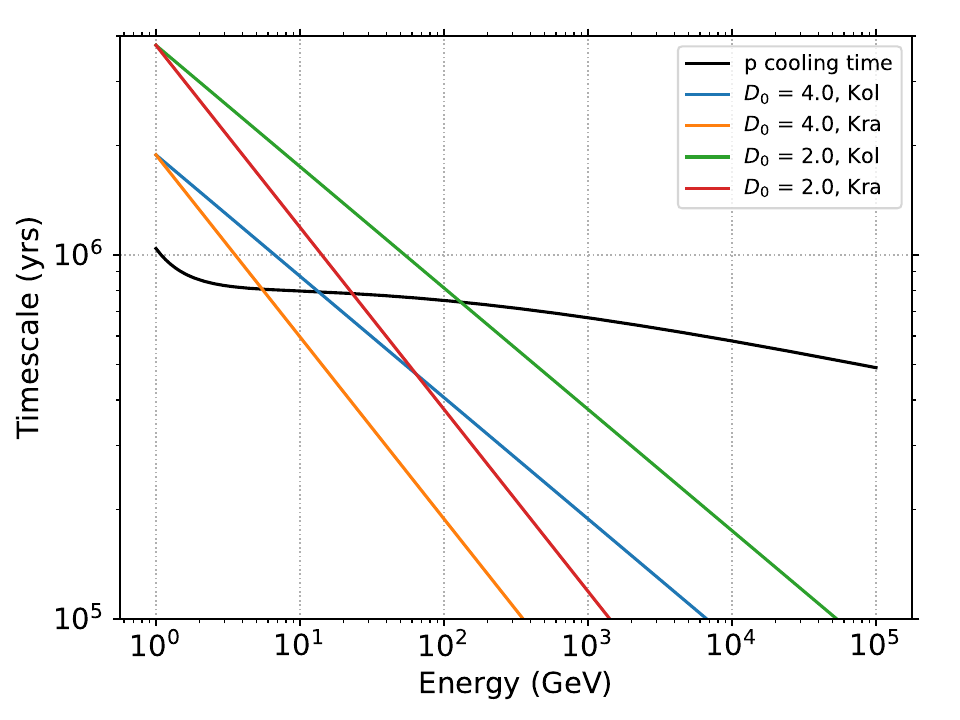}
   \caption{Timescales for proton cooling and diffusive escape near the \object{Westerlund 1} superbubble shell as a function of energy. The proton cooling time is calculated for a matter density of 150~protons~cm$^{-3}$. The colored lines provide estimates of the diffusive escape time assuming a thickness of 100~pc and either Kolmogorov ($\delta = 1/3$) or Kraichnan ($\delta=1/2$) diffusion, and for \jrone{normalizations of the diffusion coefficient at 1~GeV $D_0 = 2 \times 10^{26}$~cm$^{2}$~s$^{-1}$ or $D_0 = 4 \times 10^{26}$~cm$^{2}$~s$^{-1}$}. The upper limit of the time axis corresponds to the age of \object{Westerlund 1} of 4~Myr.
   } 
              \label{fig:ptransport}%
    \end{figure}
    
In the case of the \object{Westerlund 1} superbubble, \pbrv{\fig}~\ref{fig:ptransport} compares the diffusive escape time to cross the region of \leftblob and \centerblob with an estimated thickness of 100~pc with the proton cooling time for a matter density of 150~cm$^{-3}$. \jrone{It shows that, for diffusion coefficients with a normalization at 1 GeV $D_0 = 2\;\mathrm{to}\;4 \times 10^{26}$~cm$^{-2}$~s$^{-1}$, a transition occurs near the energy of 50 GeV where LAT data suggest a spectral drop}. In this configuration, below the transition energy protons loose all their energy interacting with the gas, while above the transition energy they escape before radiating \jrone{significantly}. The value of the diffusion coefficient is smaller by two orders of magnitudes than the average in the ISM. Regions of reduced diffusivity are expected and supported by observations around CR sources including star-forming regions \citep[e.g.,][]{tibaldoGammaRays2021, marcowith2025}. However, it remains to be established whether this can be achieved in the neutral gas region near the edge of \object{Westerlund~1} superbubble.

\section{Other potential sources of extended gamma-ray emission in the direction of \leftblob and \centerblob}\label{app:shell_alt}

Known and potential classes of objects powering extended gamma-ray emission in the LAT energy band are: SNRs, pulsars, powering pulsar wind nebulae (PWNe) and halos, and massive star-forming regions.

For \centerblob the closest SNR on the sky is SNR~338.5$+$0.1 with a size of 0.15\degr \citep{greenUpdatedCatalogue3102025} at the distance of 11~kpc \citep{kothesDistanceNeutralEnvironment2007}. 
For \leftblob there are two SNRs within the 68\% containment radius \citep{greenUpdatedCatalogue3102025}: SNR~G340.4+0.4 (size $0.17\degr \times 0.11\degr$) and SNR~G340.6+0.3 (size 0.1\degr). According to the Sigma-D relation \citep{2013ApJS..204....4P} they are at distances of 9.1~kpc and 11.2~kpc, respectively. In summary, the radio SNRs are smaller and displaced with respect to the gamma-ray emission.

For \leftblob, the most energetic pulsar within the 68\% containment radius is PSR~J1650$-$4502 with a spin-down power of $10^{34}$~erg~s$^{-1}$ and a characteristic age of 376 kyr at a distance of 3.9 kpc \citep{manchesterAustraliaTelescopeNational2005}. For \centerblob the most energetic pulsar within the 68\% containment radius is PSR~B1641$-$45 with a spin-down power of $8 \times 10^{33}$~erg~s$^{-1}$ and a characteristic age of 359 kyr at a distance of 4.5 kpc. With their ages exceeding 300~kyr, these pulsars are older than those powering the populations of known gamma-ray PWNe \citep{hesspwne,fermipwne}, although the characteristic age is subject to large uncertainties. On the other hand, middle-aged pulsars are known to power extended halos of hard gamma-ray emission with typical sizes of several tens of pc \citep[e.g.,][]{lopez-cotoGammarayHaloesPulsars2022}, that in projection roughly match the sizes of \leftblob and \centerblob at the distances of the two pulsars. However, the spectra of \leftblob and \centerblob are markedly softer of those of known pulsar halos.

As for star-forming regions, there are two clusters of young stellar objects overlapping with \leftblob and \centerblob: G340.242$-$00.37 and G338.934$-$00.062 \citep{urquhartRMSSurveyGalactic2014}. Their velocities lie in the same kinematic range of Westerlund~1 and the \hi B3 shell. From their IR luminosity they appear to be much less energetic than Westerlund~1. While the first one coincides with a peak in gas densities and gamma-ray emission, there is no evidence for enhanced gamma-ray emission coincident with the second. The star cluster catalog of \citet{2020A&A...640A...1C} does not not report any MSCs overlapping \leftblob and \centerblob, the closest one being Westerlund~1 itself.

\section{Estimate of gas densities in the region of \leftblob and \centerblob}\label{app:shell_gas_dens}

We estimate gas densities from the column density maps in \pbrv{\fig}~\ref{fig:shell_maps}. For molecular gas traced by CO we consider a spherical cloud centered on the peak of the CO map with a radius of 80~pc, which, for the standard CO-to-H$_2$ ratio $2\times 10^{20}$~molecules~cm$^{-2}$~(K~km~s$^{-1}$)$^{-1}$, yields a density of 50 hydrogen molecules cm$^{-3}$.

For the geometry of the atomic hydrogen in the template of \pbrv{\fig}~\ref{fig:shell_maps} we take projected sizes $\Delta l = 2\degr$ ($\sim$150~pc at 4.2~kpc) and $\Delta b=0.75\degr$ ($\sim$50~pc). For depths along the line of sight comprised between 150~pc and 50~pc, we obtain a density between 2 and 6 hydrogen atoms cm$^{-3}$. 

For a mean atomic weight of 1.36 in the ISM we obtain a total proton density $\sim$150 cm$^{3}$. Given the many uncertainties due, among others, to projection effects and to a varying mix of atomic and molecular gas across the regions of \leftblob and \centerblob, this shall be considered just as an indicative order-of-magnitude estimate. It applies mainly to the region of the brighter \leftblob where CO and \hi\ overlap, but considering a varying gas density is a modeling refinement beyond the scope of our paper and left for future studies. 

\end{appendix}

\end{document}